\documentclass[journal]{IEEEtran}
\usepackage{algorithmic}
\usepackage{amsmath}
\usepackage{amssymb}
\usepackage{amsfonts}
\usepackage{cite}
\usepackage{comment}
\usepackage{color} 
\usepackage{dblfloatfix}
\usepackage[T1]{fontenc}
\usepackage{float}
\usepackage{gensymb}
\usepackage{graphicx}
\usepackage[none]{hyphenat}
\usepackage{ifthen}
\usepackage[utf8]{inputenc}

\usepackage{lipsum}
\usepackage{mathtools}   
\usepackage{multirow}
\usepackage{mwe}
\usepackage{placeins}
\usepackage{subcaption}
\usepackage{soul}
\usepackage{svg}
\usepackage{tabularx,colortbl}
\usepackage{textcomp}
\usepackage[normalem]{ulem}
\usepackage{xcolor}	
\usepackage{balance}

\newcolumntype{C}{>{\centering\arraybackslash}X}

\newcommand\hlon

\begin{document}

\title{Spatial-Spectral Modeling of the Array Pattern of a Two-Element Dynamic Antenna Array with Differential Amplitude Modulation}

\author{Jacob R. Randall,~\IEEEmembership{Graduate~Student~Member,~IEEE,} Cory Hilton,~\IEEEmembership{Graduate~Student~Member,~IEEE,} \\and Jeffrey A. Nanzer,~\IEEEmembership{Senior Member,~IEEE}%
\thanks{This work was supported in part by the National Science Foundation under Grants \#2028736, \#2225337, and \#2534114. \textit{(Corresponding author: Jeffrey A. Nanzer)}}
	\thanks{	
	The authors are with the Department of Electrical and Computer Engineering, Michigan State University, East Lansing, MI 48824 USA (email: randa130@msu.edu, hiltonc2@msu.edu, nanzer@msu.edu).}
}

        %
\maketitle

\date{} 

\begin{abstract} 
We present a theoretical model for a two-element dynamic phased array and characterize the transfer of information as a function of angle. The array is based on a two-state switched structure with phase shifting to support beamsteering. Dynamic motion of the phase center of antenna arrays generates time-varying radiation patterns that, when appropriately designed, support directional modulation, or the transfer of information to regions of space that are narrower than that covered by the energy radiated by the array. We evaluate the impact of switching frequency and steering on the spatial width of the information beam, which is the region of space where information is recoverable. The concepts are evaluated through simulation and experiment using a 0.75$\lambda$ two-element array operating at 2.5 GHz.


\end{abstract}


\section{Introduction}


As the number of network devices grows exponentially, security is becoming a more critical aspect of wireless communications, increasing the likelihood of both intentional and unintentional interference between systems. 
While traditional security methods, such as baseband information encryption, remain essential, there is a growing focus on incorporating additional security measures at the physical layer. 
Physical layer security techniques have the potential to be applied independently of the underlying information, offering a method that remains transparent to the overall communication system~\cite{7258313,7120012,10848393,8335290}. 
Approaches leveraging the antenna are of particular interest, since they have the potential for implementation in both legacy and future wireless systems.
Specifically, when implemented at the aperture, physical layer security can be angle-dependent, ensuring unique signal encryption at each angle. Directional modulation (DM) achieves this by dynamically altering the antenna's radiation pattern in amplitude and/or phase outside the desired spatial region commonly referred to as the information beam~\cite{10286341,6544472,10639091,Zhang2019PSA}. As a result, while energy may be transmitted over a broad angular range, the recoverable information is restricted to a narrow beam, which can be much smaller than the half-power beamwidth (HPBW)~\cite{4684619}. 

Prior work on DM has investigated the use of amplitude and phase modulation of the transmitted signals~\cite{5422702,5159486}, or through algorithmic approaches to synthesize complex weighting factors to each radiator~\cite{6645431}. The first demonstration of DM on a single element relied on rapidly switching the current distribution across a $1.5\lambda$ dipole~\cite{10161710}. The expansion to a multi-carrier communication system has also been studied by using single-pole-triple-throw (SP3T) switches~\cite{9345966} where two states operate as a 1-bit phase shifter and the third state can turn OFF the radiator which is a traditional ON/OFF keying (OOK) technique to transmit DM~\cite{52647,6546230,8443404}. Compact DM solutions combining the complex weighting technique with a single-pole-five-throw (SP5T) can synthesize a $360\degree$ steerable information beam~\cite{9674846}, utilize stack-able multi-port patches with four unique modes combined with complex weighting~\cite{9140321}, or for a preferred single-port omnidirectional pattern, a dynamic balun~\cite{11030223}.
An alternative approach explored in our previous work is to dynamically alter the phase center of an antenna or array to rapidly alter the complex radiation pattern at angles outside of a desired information beam. We formulated an empirical model to determine the information beamwidth which is proportional to signal-to-noise ratio (SNR), element spacing, and the amplitude ratio factor~\cite{10286341}. We also have demonstrated wireless communication security of a pseudo-randomly generated bit stream of 4\;MBit/s through use of an asymmetrical Wilkinson power divider and a double-pole double-throw (DPDT) radio frequency (RF) switch~\cite{10236567}. 


In this paper, we develop a theoretical model of the radiation pattern of a two-element dynamic phased array as a function of phase center dynamics and steering of the information beam. DM capabilities are generally dependent on the spatio-temporal variation of the antenna radiation pattern, thus we characterize the radiation pattern via the spatial-angle derivative of the dynamic radiation pattern of a two-element array with relative time-varying amplitude variation. The relative amplitude between the elements is switched between two discrete states with a rapid switching frequency, which results in modulating the information into frequency sidebands, the specifics of which are dependent on the switching frequency and signal bandwidth. Using the model, we characterize the information beam as a function of steering angle for different switching frequencies and receiver bandwidths, showing that as the steering angle moves off broadside, the half-power beamwidth and information beam both increase. We compare the model to results from an experimental 0.75$\lambda$ two-element array operating at 2.5 GHz.

\begin{figure}[t!]
\centering
\includegraphics[width=0.9\columnwidth]{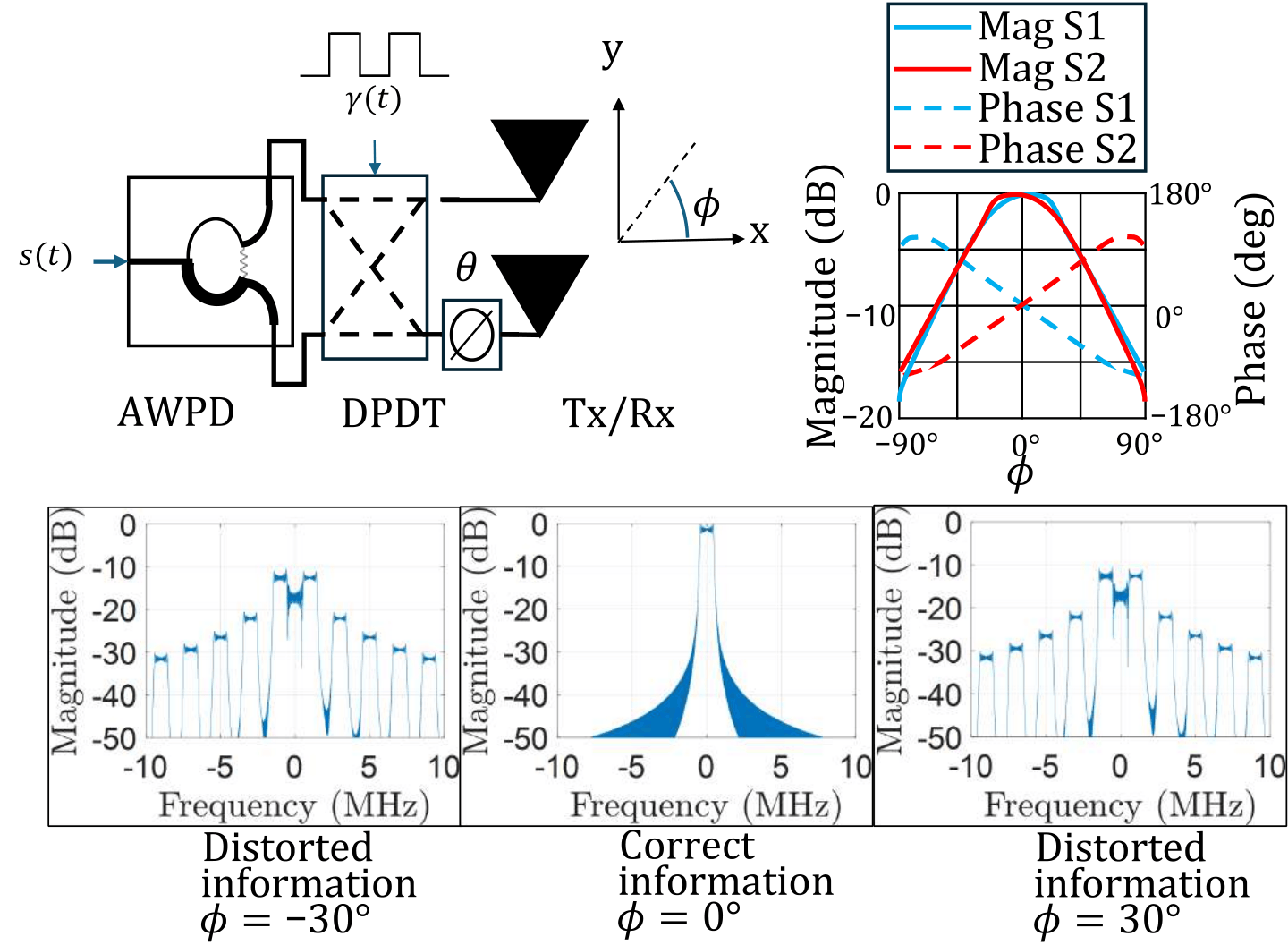}
\caption{The dynamic antenna concept is based on an asymmetric Wilkinson Power Divider (AWPD), a double-pole double-throw (DPDT) radio frequency (RF) switch, and an RF phase shifter. Rapidly switching between the two feeding states dynamically moves the antenna phase center. The resultant radiation pattern in have similar magnitude but the phase front is odd symmetric about broadside, inducing modulaton on signals transmitted or received outside of the broadside direction (this direction can be steered with the phase shifter).}
\label{fig:OV1}
\end{figure}

\section{Dynamic Antenna Theory}

The dynamic antenna concept is illustrated in Fig.~\ref{fig:OV1}. The antenna consists of a two-element array with an asymmetric power divider that splits the input signal into two paths with unequal power. The signals pass through a DPDT switch which determines which of the two antennas is fed by the output of the power divider. By rapidly switching the DPDT, two antenna states are generated, where the power imbalance causes the phase center to move, resulting in two unique complex radiation patterns. The goal is to generate a pattern with a magnitude that is constant between the two states, but with phase patterns that are antisymmetric around the direction that secure information will be transmitted. The differences in phase between the two states creates an additional modulation on signals transmitted or received outside of the intended direction, corrupting the data and increasing the bit errors. Essentially, the fast switching modulates the information at angles away from the intended direction, making the information more challenging to recover.


We characterize the dynamic array factor of a two-element dynamic phased array as
\begin{align}
AF(t,\phi,\theta)& \nonumber \\ = E(\phi)&\left\{\gamma(t)e^{-jk\frac{d}{2}\sin\phi}e^{j\theta} + [1-\gamma(t)]e^{jk\frac{d}{2}\sin\phi}\right\}
\label{eqn:AF}
\end{align}
where $E(\phi)$ is the far-field radiation pattern of a single element (assumed to be identical for the two antennas), $\gamma(t)$ is the time-varying amplitude distribution factor which has the constraint $0\leq\gamma(t)\leq1$, $k$ is the wavenumber, $d$ is the element spacing, and $\theta$ is a relative phase offset between the two element excitations. 

The time-varying phase center of the array can be given by~\cite{10286341}
\begin{equation}\label{}
x_{pc}(t) = \frac{d}{2}\left[1 - 2\gamma(t)\right]
\label{eqn:x_pc}
\end{equation}
In practical implementations, the phase center is dynamically varied between two discrete locations, which is obtained by asymmetrically splitting the signal power to the two elements, and switching the feeds rapidly such that the amplitude ratio is imparted in two mirrored states. 



The magnitude and phase of the array factor are given by
\begin{align}\label{magAF}
  &\left|AF\left(t,\phi,\theta\right)\right| = \nonumber \\  &\sqrt{\gamma^2(t) + \left[1-\gamma(t)\right]^2 + 2\gamma(t) \left[1-\gamma(t)\right] \cos\left[kd\cos\left(\phi-\theta\right)\right]}
\end{align}
and
\begin{equation}\label{phaseAF1}
\angle AF(t,\phi,\theta) = \tan^{-1}\left(\frac{A}{B}\right)
\end{equation}
where
\begin{equation}\label{eqn:phaseAF}
\frac{A}{B} = \frac{\left[1-\gamma(t)\right]\sin[\frac{kd}{2}\cos\phi]-\gamma(t)\sin[\frac{kd}{2}\cos(\phi-\theta)]}{\left[1-\gamma(t)\right]\cos[\frac{kd}{2}\cos\phi]+\gamma(t)\cos[\frac{kd}{2}\cos(\phi-\theta)]}
\end{equation}
From these, it can be seen that the magnitude is an even function of the angle $\phi-\theta$. For a system switching between two mirrored amplitude feeds to the elements, the magnitude of the array pattern remains the same due to the symmetry around $\phi-\theta$. The phase pattern, however, is anti-symmetric about $\phi-\theta$, thus as the feeds are rapidly switched the phase pattern switches between two different states, creating a time-varying phase modulation that changes as a function of angle. At $\phi-\theta$ the phase difference is zero, therefore the angles where no dynamics occur can be given as
\begin{equation}
\phi = \sin^{-1}\left[\frac{m\pi+2\theta}{kd}\right], \;m=[...,-2,0,2,...]
\label{eqn:FundamentalPower_MaxAngle}
\end{equation}
Conversely, the angles where the maximum dynamics occur is the angle where the two radiators add destructively, given as
\begin{equation}
 \phi = \sin^{-1}\left[\frac{n\pi+2\theta}{kd}\right], \;n=[...,-3,-1,1,3,...]
\label{eqn:HarmonicPower_MaxAngle}
\end{equation}

The time-varying phase center moves with a profile proportional to $\gamma(t)$. In this work we implement a two-state discrete square wave function, given by
\begin{equation}
\gamma(t) = \frac{4\alpha}{\pi}\sum_{n=1,2,...}^{\infty}\frac{1}{2n-1}\mathrm{sin}\left[\frac{(2n-1)\pi t}{T_0}\right]
\label{eqn:PC_Motion_Squarewave}
\end{equation}
where the periodicity of the spatial amplitude dynamics is $T_0$ and the amplitude $0 \leq \alpha \leq 1$ is a scaling term that determines the range over which the phase center moves; $\alpha = 0$ indicates no phase center movement while $\alpha = 1$ indicates movement over the full range of the aperture. This amplitude can be set by the relative amplitudes of the signals fed to the two antennas, which are then rapidly switched, e.g., identical signal amplitudes yield $\alpha = 0$.

A larger amplitude imbalance $\mathrm{max}[\gamma(t)]$ will result in stronger modulation products due to the phase center moving a larger distance as denoted in \eqref{eqn:x_pc}. Increasing the element spacing $d$ will also increase how far the phase center moves; however due to \eqref{eqn:FundamentalPower_MaxAngle}, large element spacings result in more than one angular region where the phase center motion appears purely tangential due to the phase center moving a distance equal to an an electrical length of one wavelength. Analogous to traditional array theory grating lobes, designs may benefit from keeping the element spacing under $1\lambda$ and the relative phase offset for beamforming $-kd~<~\theta~<~kd$. 

Because the array pattern is time-dependent, information transmitted or received by the dynamic array may be modulated into frequency sidebands in areas where the apparent phase center movement is large. This effect can be characterized by analyzing the spatial-spectral response of the array pattern, given by 
\begin{equation}
\tilde{P}(\omega,\phi,\theta) = \mathcal{F}\left\{{AF}(t,\phi,\theta)\right\}
\end{equation}
where $\mathcal{F}\left\{\cdot\right\}$ indicates Fourier transform. While this characterizes the response in general, it is helpful to explore simpler models of how the information is distributed in space and time. To do this, we characterize a form of the time-varying array factor that separates the components existing at the fundamental frequency from those being modulated into adjacent bands. In this heuristic model the array factor is given by
\begin{equation}
    \begin{aligned}
        & AF_T(t,\phi,\theta) = \overline{AF}(t,\phi,\theta) + \gamma(t)\frac{\partial}{\partial\phi}\{\overline{AF}(t,\phi,\theta)\}
    \end{aligned}
\label{eqn:Final_AF}
\end{equation}
The first term of~\eqref{eqn:Final_AF} captures the effect of the spatial distribution of the information at the fundamental frequency, where the average array factor is given by
\begin{equation}
\overline{AF}(t,\phi,\theta) = E(\phi)\overline{\gamma}(t)\left[e^{-j\frac{kd}{2}\mathrm{sin}\phi}+e^{j\theta}e^{j\frac{kd}{2}\mathrm{sin}\phi}\right]
\label{eqn:TimeAveraged_ArrayFactor}
\end{equation}
where the average distribution factor $\overline{\gamma}(t)$ is 0.5 for mirrored switching. The second term of~\eqref{eqn:Final_AF} captures the effect of the information being modulated into frequency sidebands as a function of angle. To model this effect, we consider a spatial power distribution calculated by the spatial derivative of the array factor yielding

\begin{align}
\label{eqn:Harmonic_Power_Theory1}
\frac{\partial}{\partial\phi}\overline{AF}(t,\phi,\theta) 
&= 2\overline{\gamma}(t)\Bigg[
\cos\left(\frac{kd}{2}\sin\phi + \frac{\theta}{2}\right)\frac{\partial}{\partial\phi}E(\phi) \nonumber \\
&\quad + \frac{kd}{2}E(\phi)\cos\phi\sin\left(\frac{kd}{2}\sin\phi + \frac{\theta}{2}\right)
\Bigg]
\end{align}

In the simple case where the antennas are isotropic radiators, $|E(\phi)| = 1$, and $\angle E(\phi)$ is constant, therefore $\partial E(\phi)/\partial \phi=0$ and the spatial power distribution reduces to
\begin{equation}
\frac{\partial}{\partial\phi}\overline{AF}(t,\phi,\theta) = kd\overline{\gamma}(t)\cos\phi\sin\left(\frac{kd}{2}\sin\phi + \frac{\theta}{2}\right)
\label{eqn:Harmonic_Power_Theory2}
\end{equation}
While \eqref{eqn:Harmonic_Power_Theory2} captures the spatial distribution of the energy in the frequency sidebands, its frequency distribution is determined by the time-varying function $\gamma(t)$, thus to appropriately account for the spatial and spectral dependence we multiply \eqref{eqn:Harmonic_Power_Theory2} by $\gamma(t)$, yielding the second term in~\eqref{eqn:Final_AF}. In the following, we simulate~\eqref{eqn:Final_AF} and compare it to full-wave numerical simulations and experimental results, validating the model.

\begin{figure}[t!]
\centering
\includegraphics[width=0.999\columnwidth]{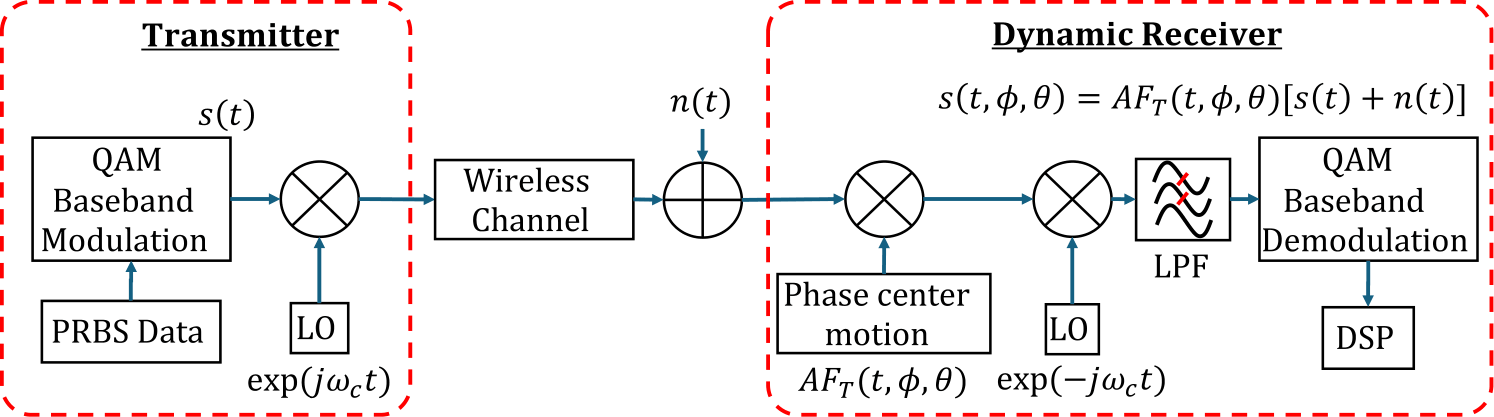}
\caption{System block diagram of the dynamic receiver due to a wireless channel in the presence of additive white Gaussian noise $n(t)$. 
}
\label{fig:Comms_System_Diagram}
\end{figure}

\subsection{Antenna Design}

To validate the dynamic phase center concept, a two-element patch antenna array was designed to resonate at 2.5\;GHz with an element spacing of $0.75\lambda$ (see~\cite{11266700,10286341}). The antenna was simulated in Ansys High Frequency Structure Simulator (HFSS) with a Rogers RO4350B substrate with a height of 1.542\;mm and permittivity of 3.66. The length and width of the patches were 2.995\;cm and 3.928\;cm, respectively. The length and width of the ground plane and substrate were 18.1\;cm and 9.95\;cm, respectively. The feed point was designed to be 0.96\;cm from the patch edge to impedance match the antenna. The simulated gain was 5.74\;dBi; a fabricated single element yielded a measured gain of 5.7\;dBi in a fully enclosed anechoic chamber environment at 2.5\;GHz.
The two-element array was then constructed and evaluated in HFSS with an asymmetric amplitude balance to each radiator. The radiation pattern of the single element was imported into MATLAB to synthesize the dynamic array radiation pattern using~\eqref{eqn:AF}. Two 2.5\;GHz single element antennas were manufactured and placed physically $0.75\lambda$ apart.

The antenna array was integrated into the dynamic feed structure shown in Fig. \ref{fig:OV1}, consisting of a phase shifter, used to calibrate phase differences of the two channels and for steering the information beam, and an asymmetrical Wilkinson divider designed with an S21 of -1.08\;dB and S31 of -7.23\;dB, resulting in a 6.15\;dB power ratio~\cite{10236567}. Between the feed structure and the antenna aperture a double-pole-double-throw RF switch was placed to rapidly switch the power balance between two mirrored states on the two elements. The directionally modulated fields generated by this dynamic antenna aperture are shown in Fig.~\ref{fig:OV1}. The magnitudes of the patterns of the two states are effectively identical, whereas the phase patterns are anti-symmetric and different between the two states at all angles except at broadside. These phase differences, when implemented in rapid switching between the two states, imparts modulation onto the signals at all angles except where the phase difference is zero, therefore creating a narrow information beam at that angle. While the angle shown here is at broadside, this angle can be steered with the a relative static phase shift between the antennas feeds. 

The impact of the directional modulation can be visualized by looking at the spatial-spectral power density of the transmitted or received signals. Since the total power is conserved, the energy at angles away from the information beam is modulated into higher frequency bands which may be impacted by the bandwidth limitations of the transceiver hardware. The signal path can be represented by the block diagram in Fig.~\ref{fig:Comms_System_Diagram}. The baseband information, $s(t)$ passes through a simple wireless channel with additive white Gaussian noise $n(t)$. Assuming the wireless channel only contains the line-of-sight (LoS) with no amplitude or phase distortion and the antenna aperture has sufficient bandwidth, the directionally modulated received signal can be represented given as 
\begin{equation}
    \begin{aligned}
        s(t,\phi,\theta) = AF_T(t,\phi,\theta)[s(t)+n(t)]
\end{aligned}
\label{eqn:Dynamic_Received_Signal}
\end{equation}
Note that if the RF switching frequency exceeds the receiver bandwidth, the signal will undergo the effects of filtering and higher-order intermodulation products induced by $AF_T(t,\phi,\theta)$ will be minimized.

\begin{figure}[t!]
\centering
\includegraphics[width=0.6\columnwidth]{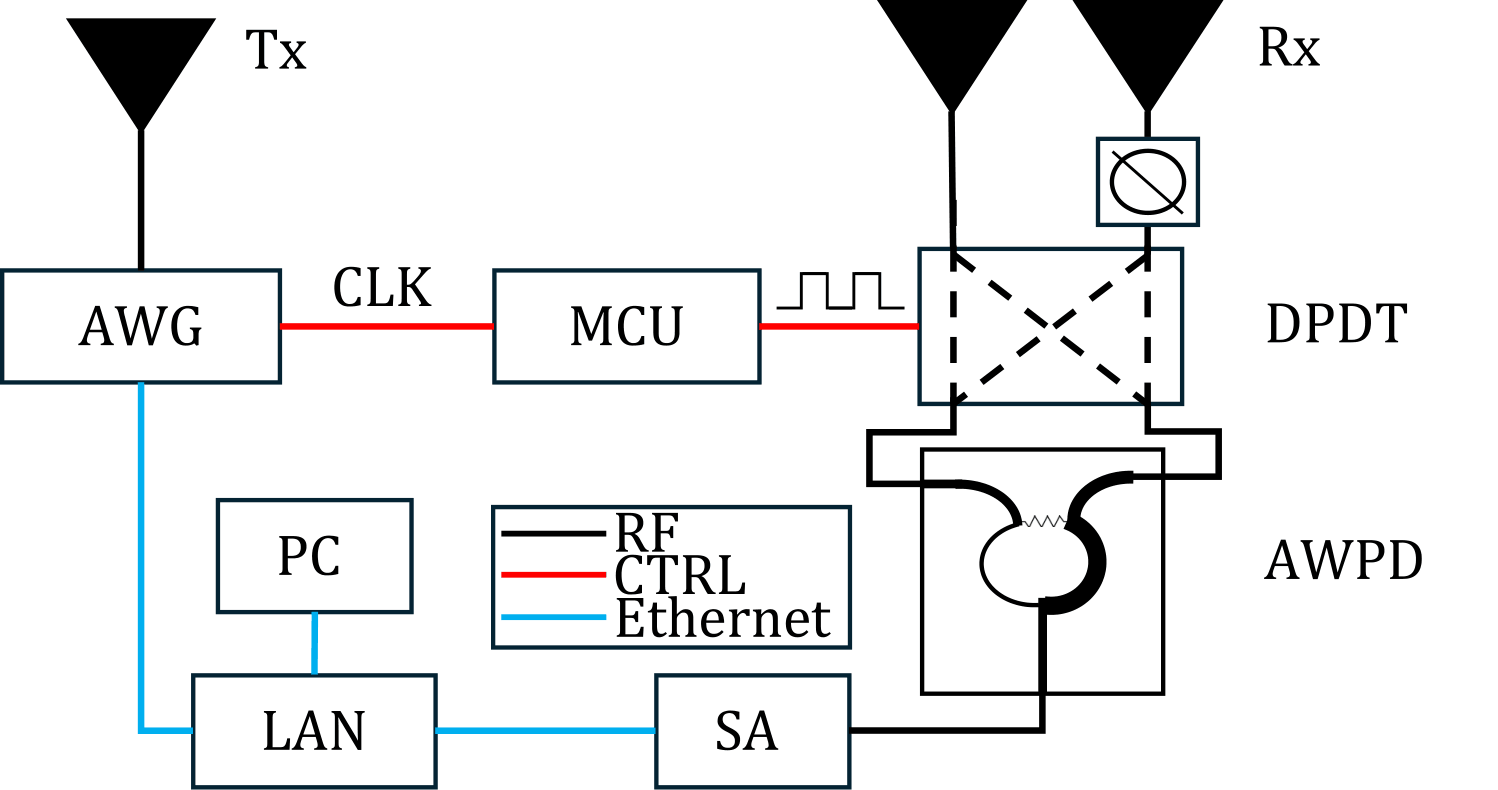}\\(a)\\
\centering
\includegraphics[width=0.6\columnwidth]{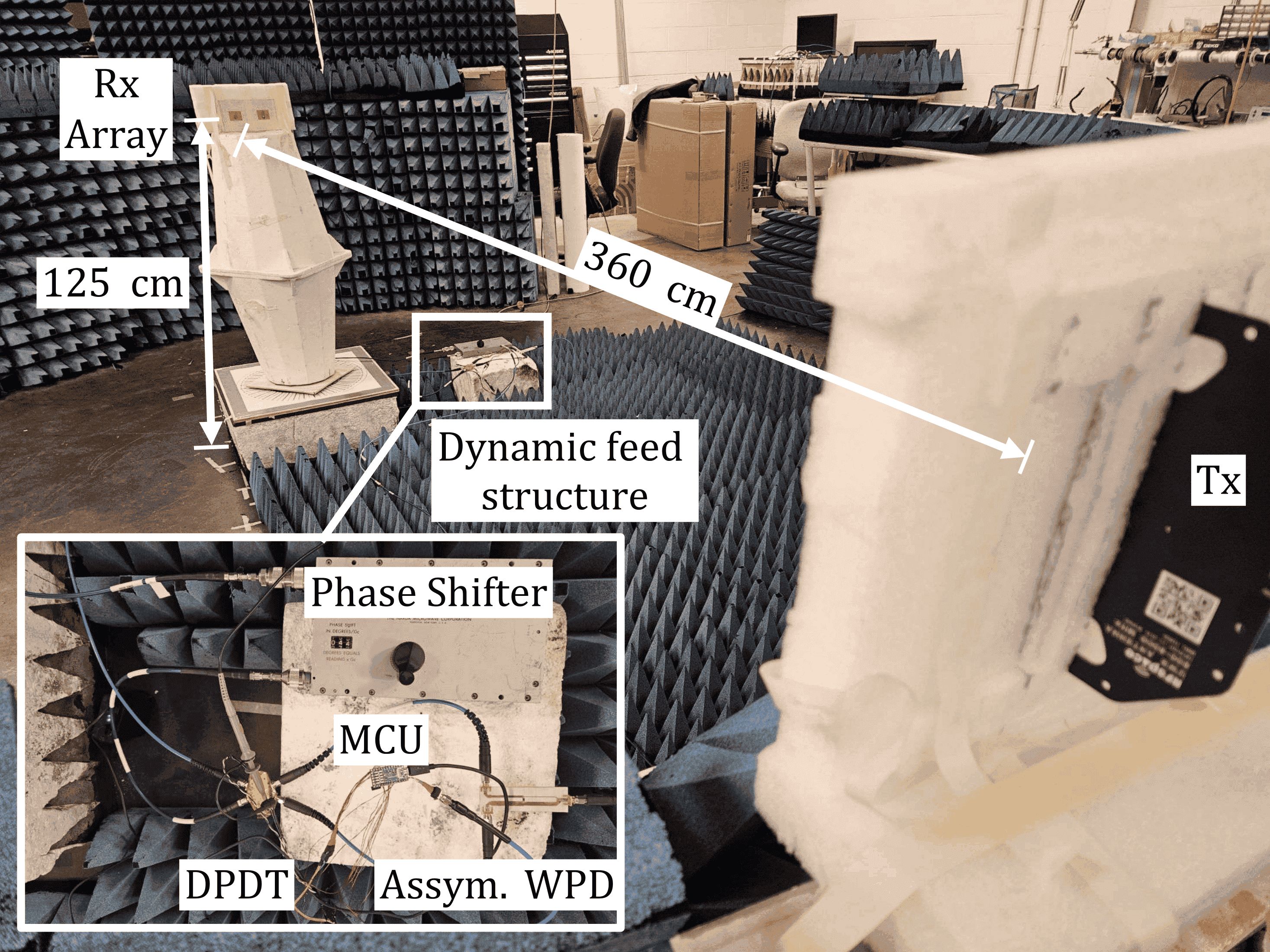}\\(b)\\
\caption{Measurement design of the dynamic phased array (a) and experimental setup in a semi-anechoic environment (b).}
\label{fig:Measurement_Setup}
\end{figure}

\begin{figure}[t!]
\centering
\includegraphics[width=0.8\columnwidth]{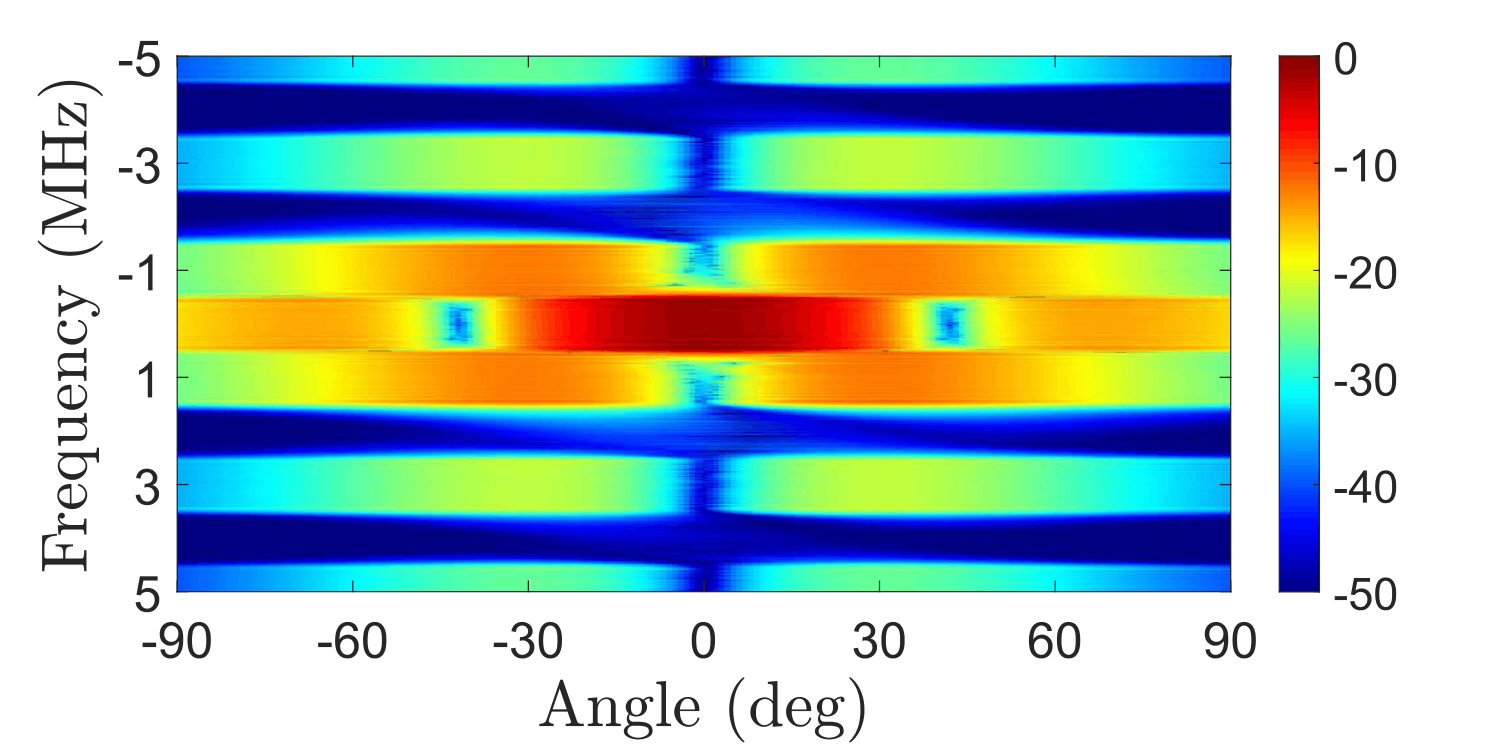}\\(a)\\
\includegraphics[width=0.8\columnwidth]{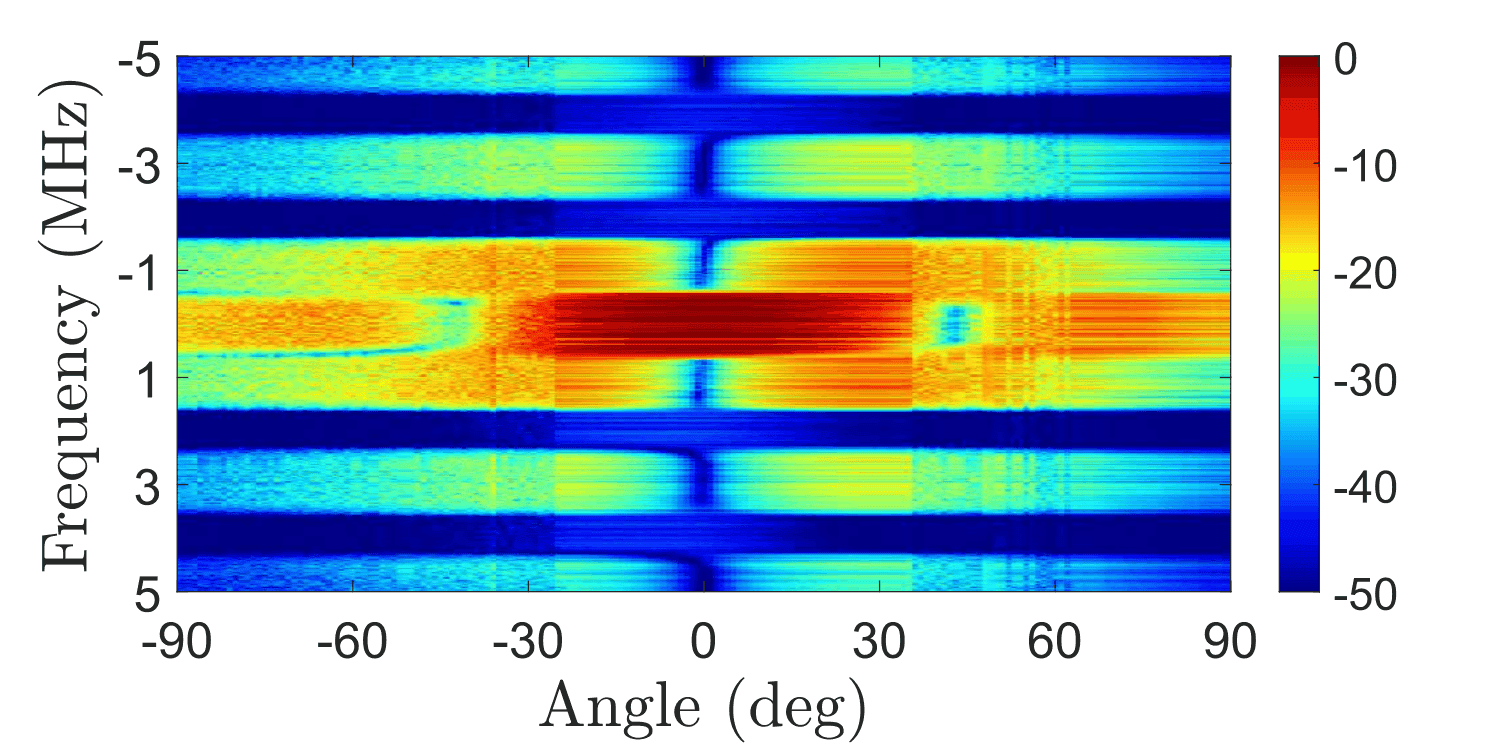}\\(b)\\
\caption{Spatial-spectral power density with a two-element patch antenna array in simulation (a) and measurement (b).
}
\label{fig:Dynamic_Antenna_PSD}
\end{figure}

\begin{figure}[t!]
\centering
\includegraphics[width=0.49\columnwidth]{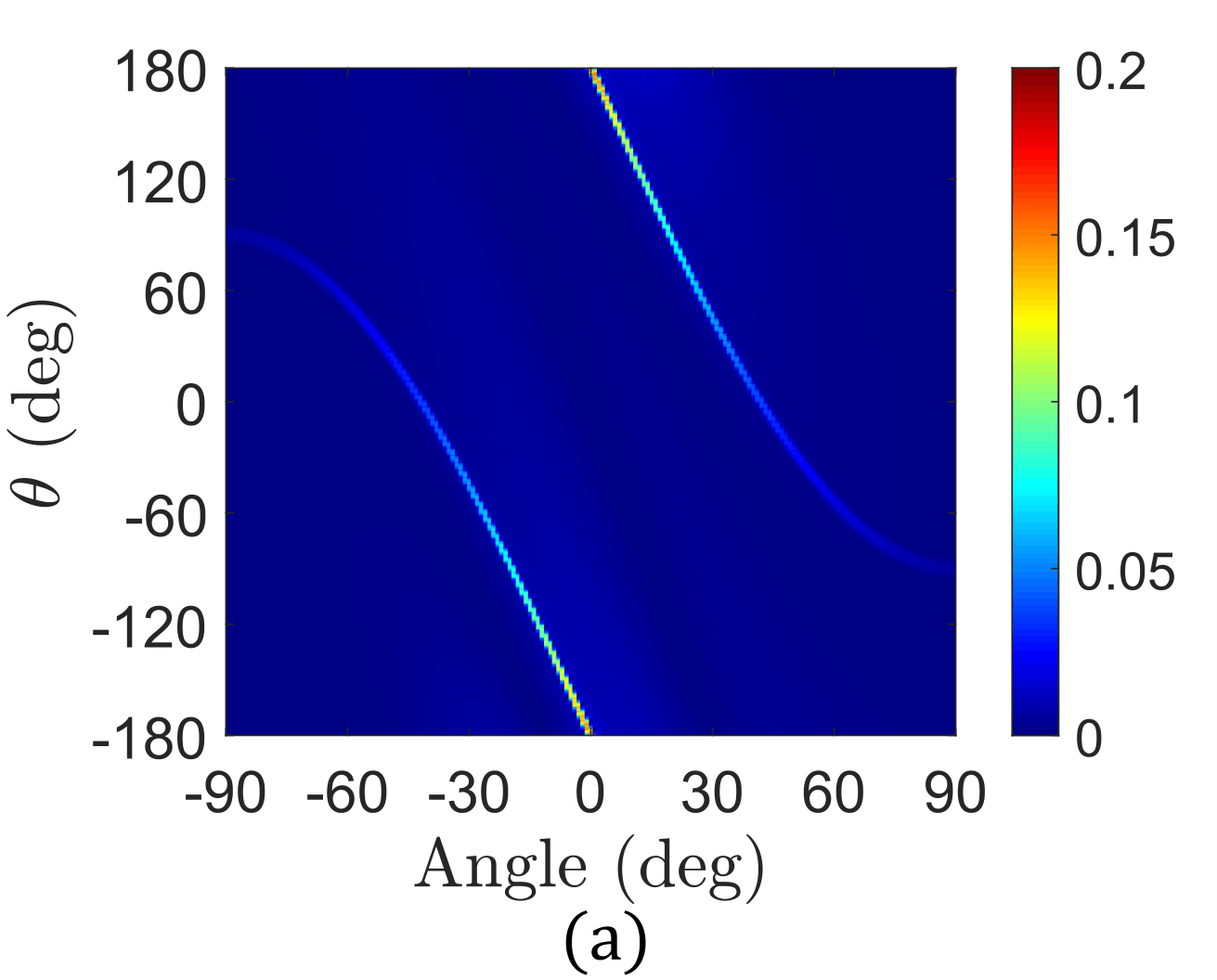}
\includegraphics[width=0.49\columnwidth]{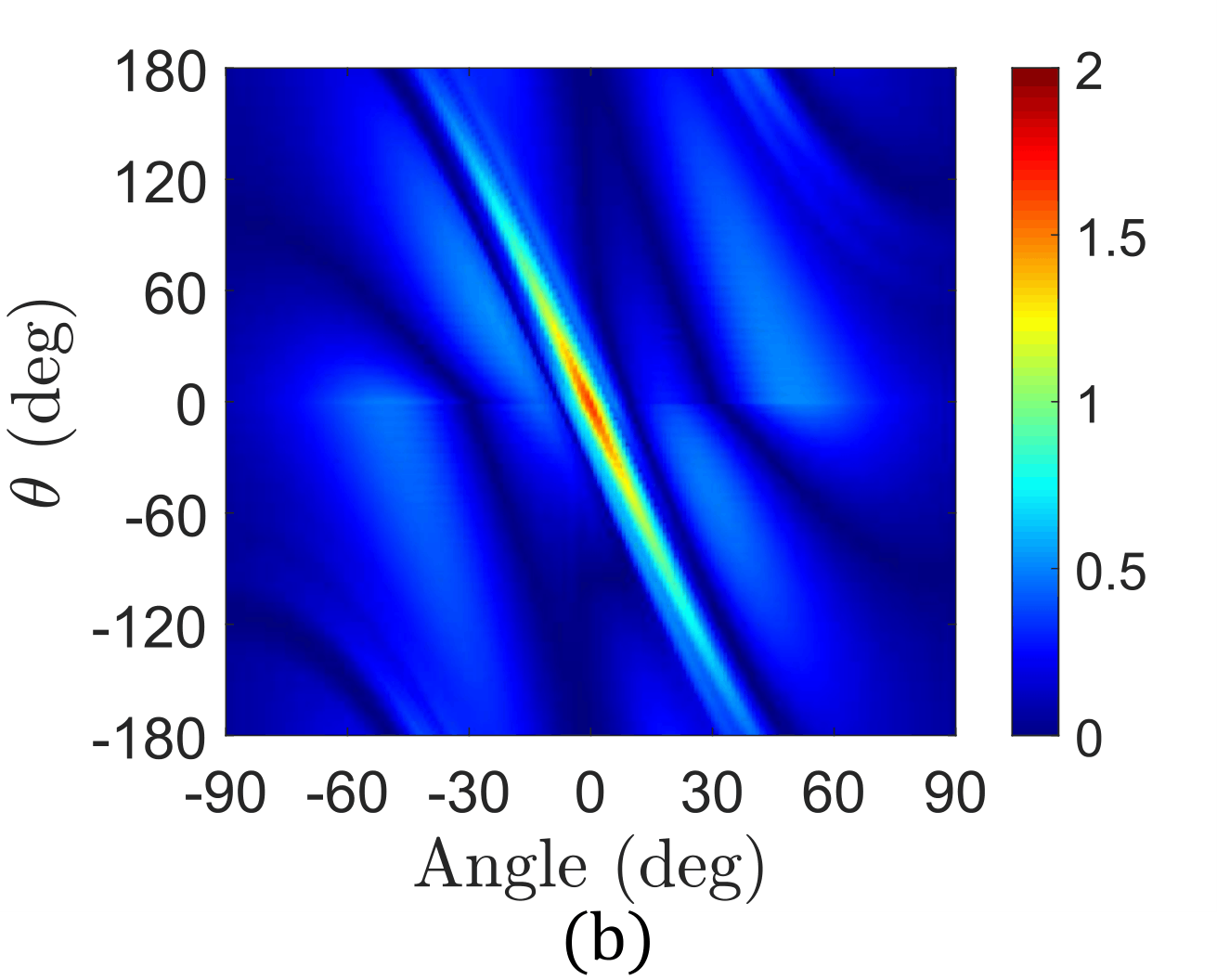}

\caption{Absolute error ($\%$) of the simulated PSD relative to the model over the full observation window $\phi$ and all relative phase offsets $\theta$ for the dynamic patch antenna array. The absolute error of the time-averaged radiation pattern (a) component specified by \eqref{eqn:TimeAveraged_ArrayFactor}. The intermodulation products with the spatial amplitude dynamics given for the patch radiator (b) using \eqref{eqn:Harmonic_Power_Theory1} is observed. The RMSE of the model across all observation and steering angles is $0.009\%$ and $0.282\%$, respectively. 
}
\label{fig:Dynamic_Antenna_Pattern_SimvsMeas}
\end{figure}

\section{Dynamic Phased Array Design}

\subsection{Dynamic Phase Center Model Validation}


We evaluate the dynamic antenna model by simulating both the general expression for the array factor~\eqref{eqn:AF} and the heuristic model~\eqref{eqn:Final_AF}, and compare both to measurements in an experimental system. 
The experimental system diagram shown in Fig.~\ref{fig:Measurement_Setup} (a) and in the far field of the antenna array shown in Fig.~\ref{fig:Measurement_Setup} (b) in a semi-anechoic environment to minimize the effects of multipath. We consider a 16-QAM signal in all cases, transmitting 48,000 bits per measurement, and the array modulation function $\gamma(t)$ was switched at a rate of 1.0\;MHz. The dynamic antenna was used as the receiver with the communications signal transmitted on a 2.5 GHz carrier frequency using an arbitrary waveform generator (Keysight M8190A) using a TSA800 ultra wideband Vivaldi antenna. The receiving array was connected to the dynamic feed structure composed of a DPDT RF switch (SKY13411-374LF-EVB), a coaxial phase shifter (Narda 3752), and an asymmetrical Wilkinson power divider (AWPD)~\cite{10236567}. 
The relative amplitudes of the two antenna channels was set to $\pm6.15$\;dB for the two dynamic antenna states at the carrier frequency of 2.5\;GHz.
The AWG was also used to generate a clock signal for the microcontroller (MCU). After passing through the AWPD, the signals were captured by a signal analyzer (Keysight N9030A) which down-converted and digitized the signals. The sampled data was then processed using the Keysight Vector Signal Analysis (VSA) 89600 software to measure the power and compute the error vector magnitude (EVM) and bit error ratio (BER).

The theoretical and heuristic models are compared to experimental measurements by analyzing the spatial power spectral density of the received signal. The theory and heuristic simulations implemented $E\left(\phi\right)$ using the HFSS-derived element pattern.
The relative phase offset $\theta$ was iteratively swept in the range $\theta \in [-\pi, \pi]$ over an observation angle $\phi \in [\frac{-\pi}{2},\frac{\pi}{2}]$. 
The spatial PSD for the model and the experiment are shown in Fig.~\ref{fig:Dynamic_Antenna_PSD}. The 1 MHz modulation of the dynamic array has a clear effect at angles away from broadside, where significant spectral sidebands manifest. At broadside, all of the spectral content is contained at the center frequency (DC for the baseband signal) because the dynamic array implemented little to no dynamic modulation at this angle. Note that at the center frequency the nulls in the pattern match those expected with the $0.75\lambda$ spacing of the antenna elements. The spectral sideband content correlates to \eqref{eqn:Harmonic_Power_Theory1} and maximizes at the null of the time-averaged radiation pattern component, correlating to \eqref{eqn:FundamentalPower_MaxAngle}. The sideband power peaks at -12.2\;dB for the patch array which is indicative of the directivity of the antenna used; for an isotropic element pattern it peaks at -9.03\;dB. These power values are governed by the spatial derivative of the radiation pattern~\eqref{eqn:Harmonic_Power_Theory1} where the theoretical maximum is $\mathrm{max}[\gamma(t)]-\frac{1}{2} \times \overline{\gamma(t)}|=0.125=-9.03$\;dB. Higher-order sideband products then follow a $1/n$ amplitude tapering, typical of square-wave excitation. The measured peak side band power of -13.8\;dB indicates good agreement between the simulated model and the experimental setup measurements.


\begin{figure}[t!]
\centering
\includegraphics[width=0.60\columnwidth]{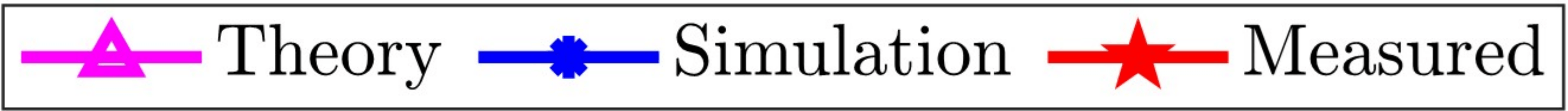}\\
\includegraphics[width=0.49\columnwidth]{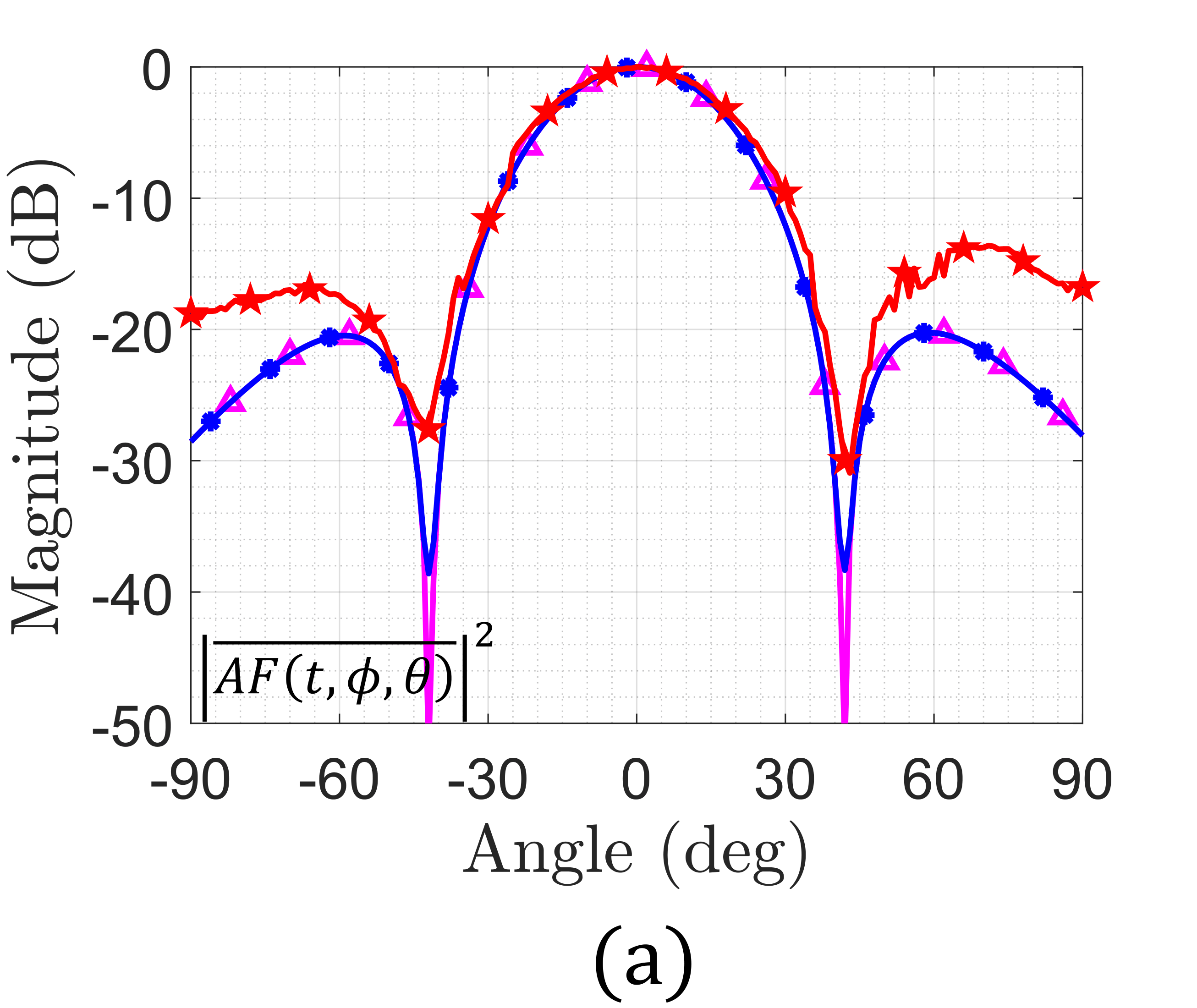}
\includegraphics[width=0.49\columnwidth]{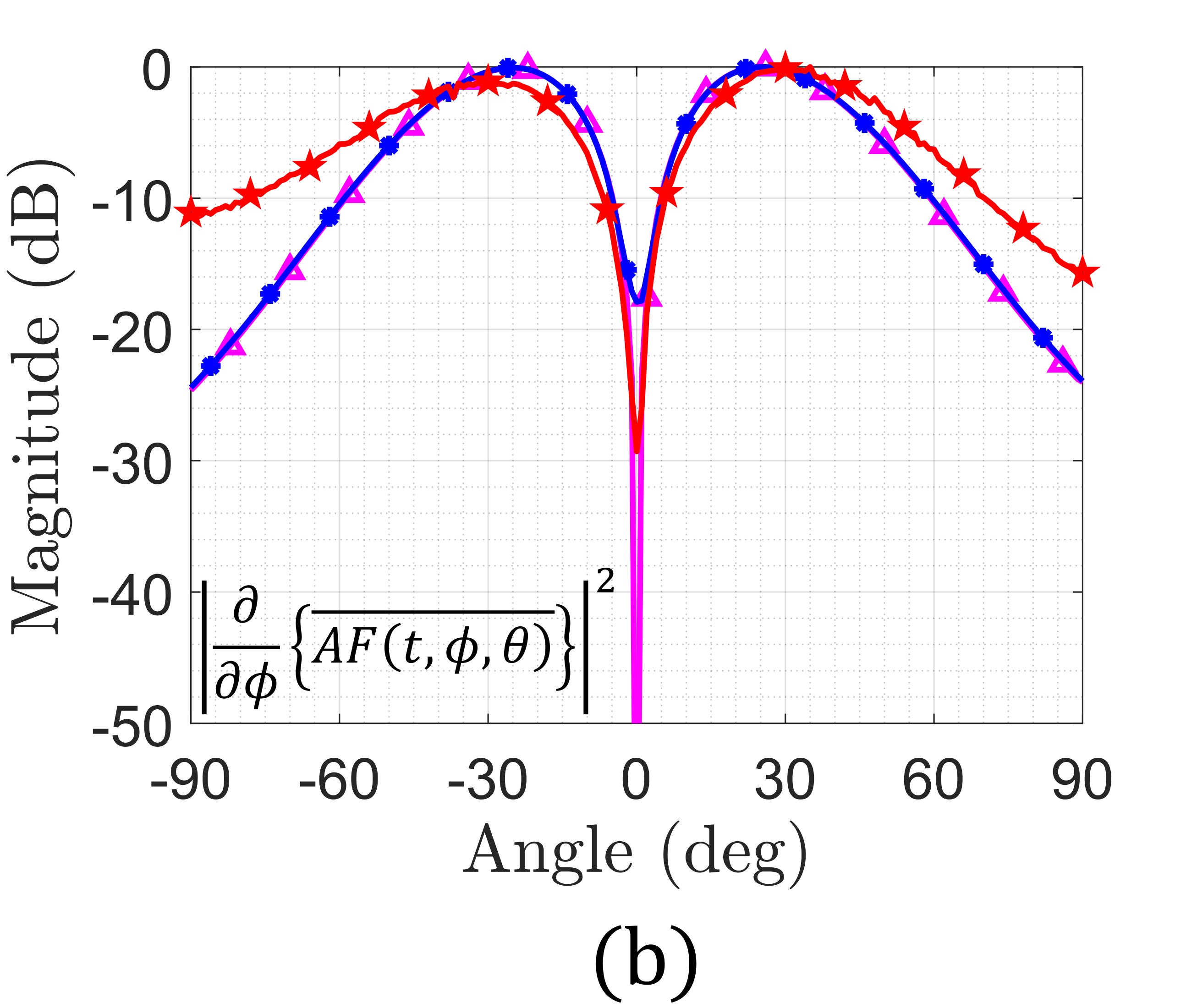}
\includegraphics[width=0.49\columnwidth]{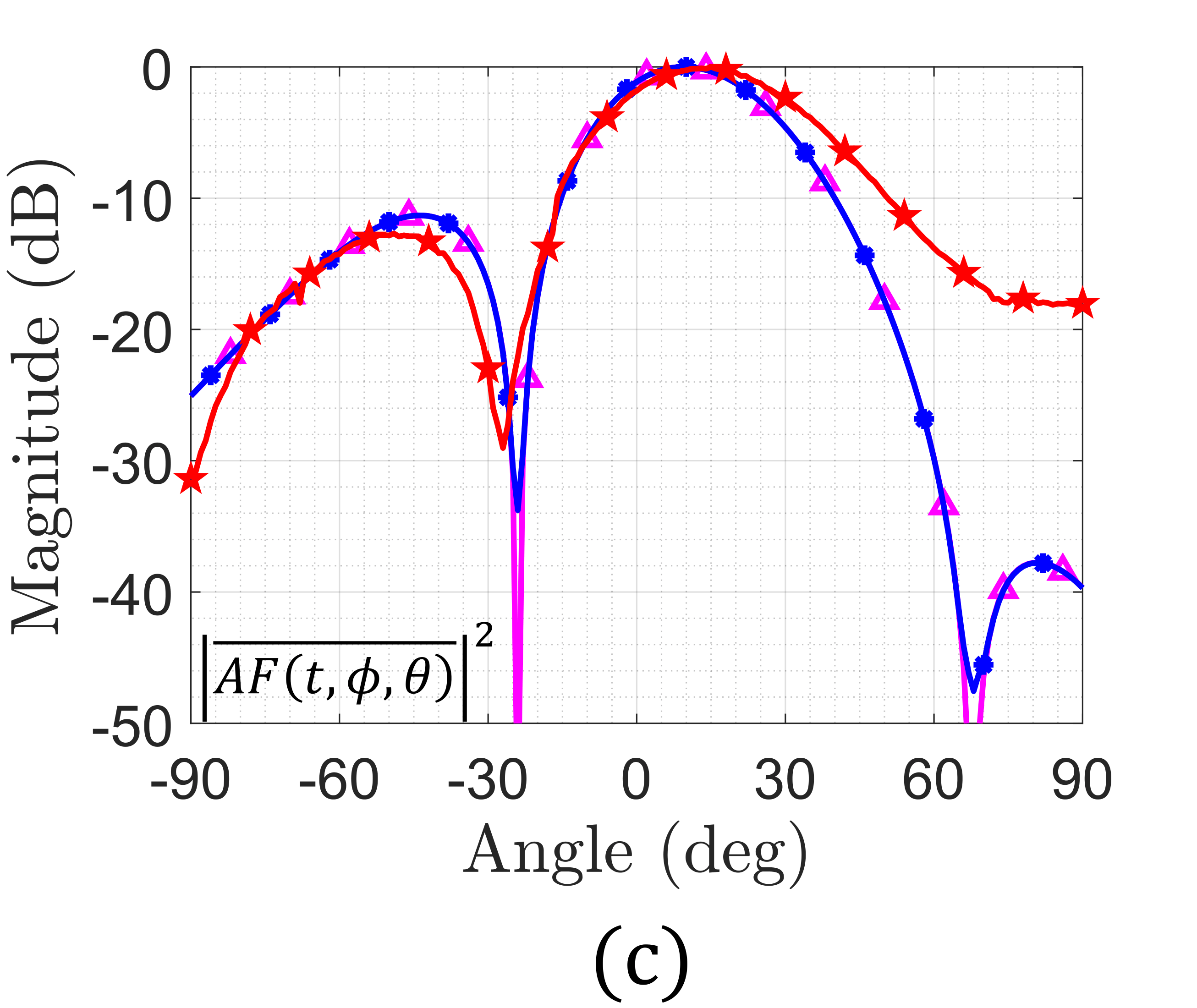}
\includegraphics[width=0.49\columnwidth]{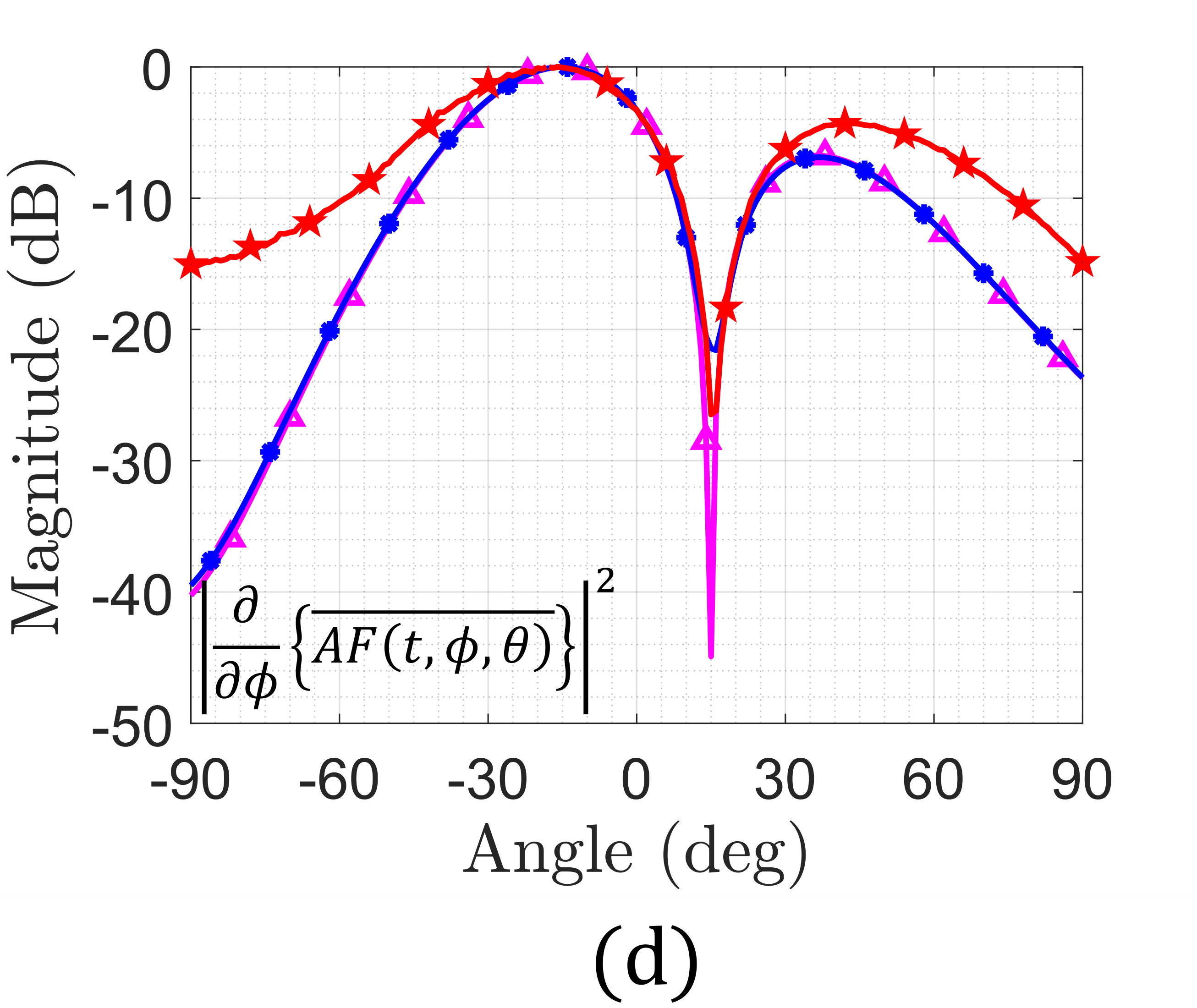}
\includegraphics[width=0.49\columnwidth]{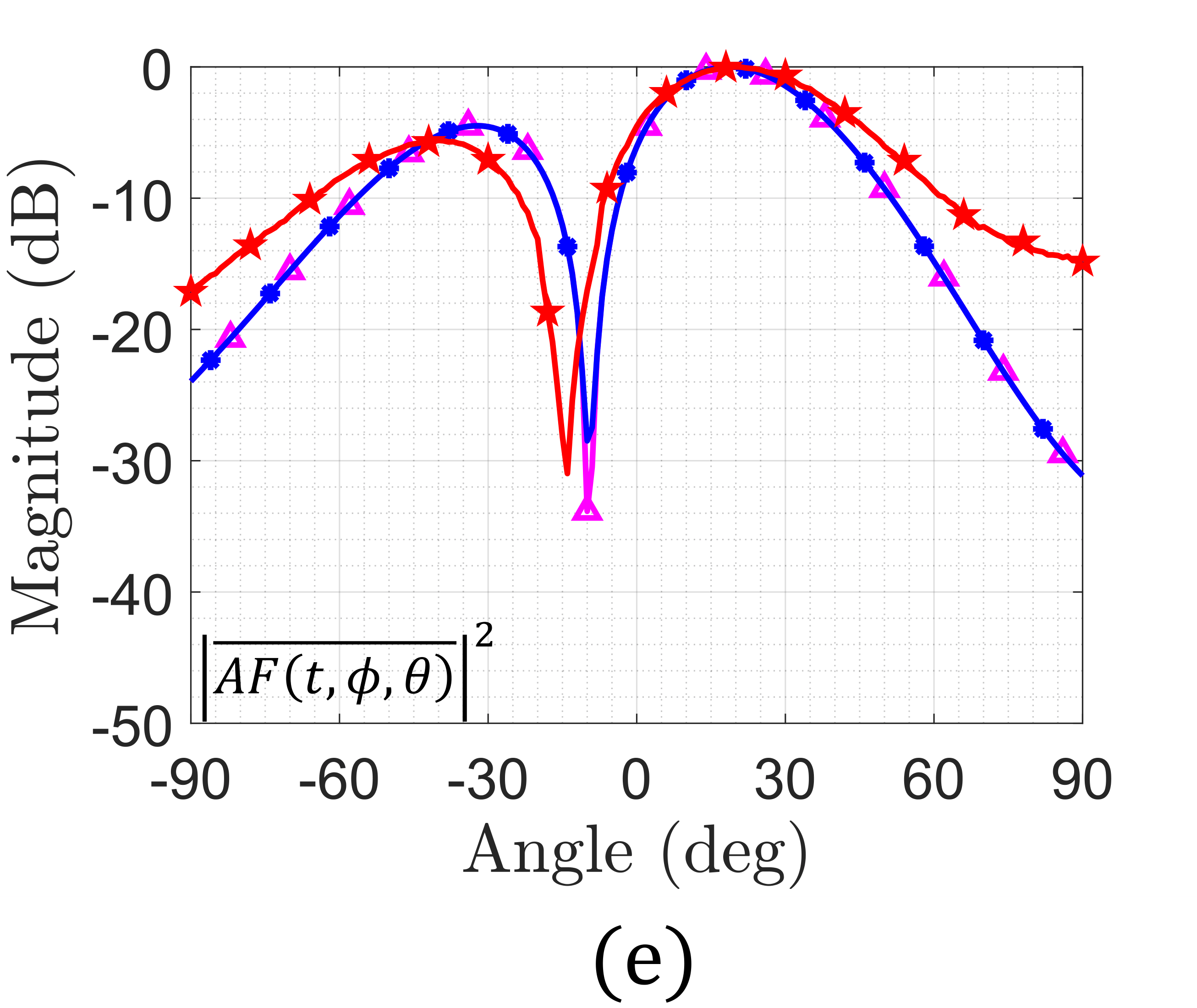}
\includegraphics[width=0.49\columnwidth]{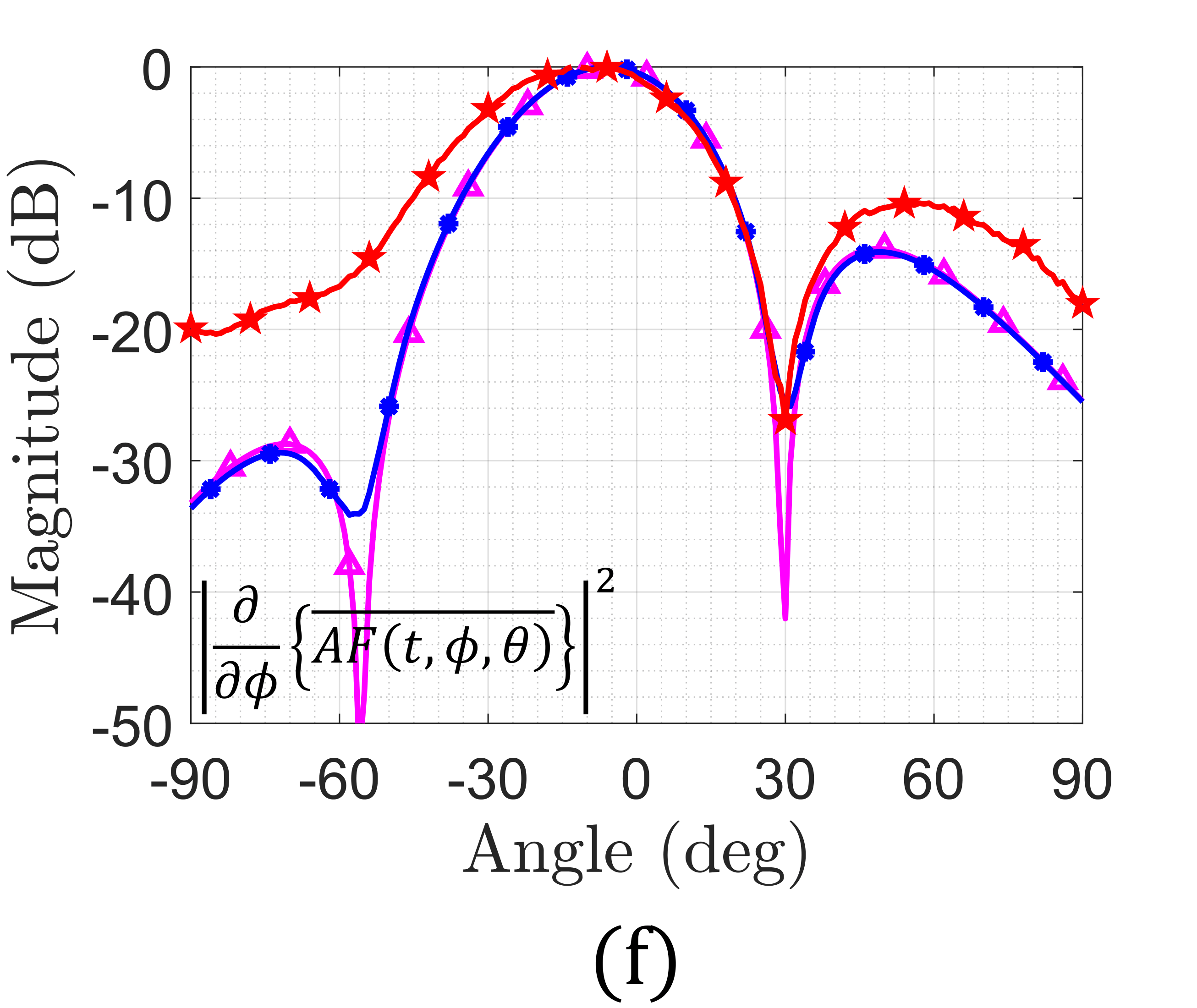}
\caption{Steering is implemented using a phase shifter. 
The measured time-averaged radiation (a), (c), (e), and intermodulation radiation (b), (d), (f), match well between the model, simulation, and measurement. 
}
\label{fig:Dynamic_Antenna_Array_Model_SimvsMeasured}
\end{figure}

\begin{figure}[t!]
\centering
\includegraphics[width=0.99\columnwidth]{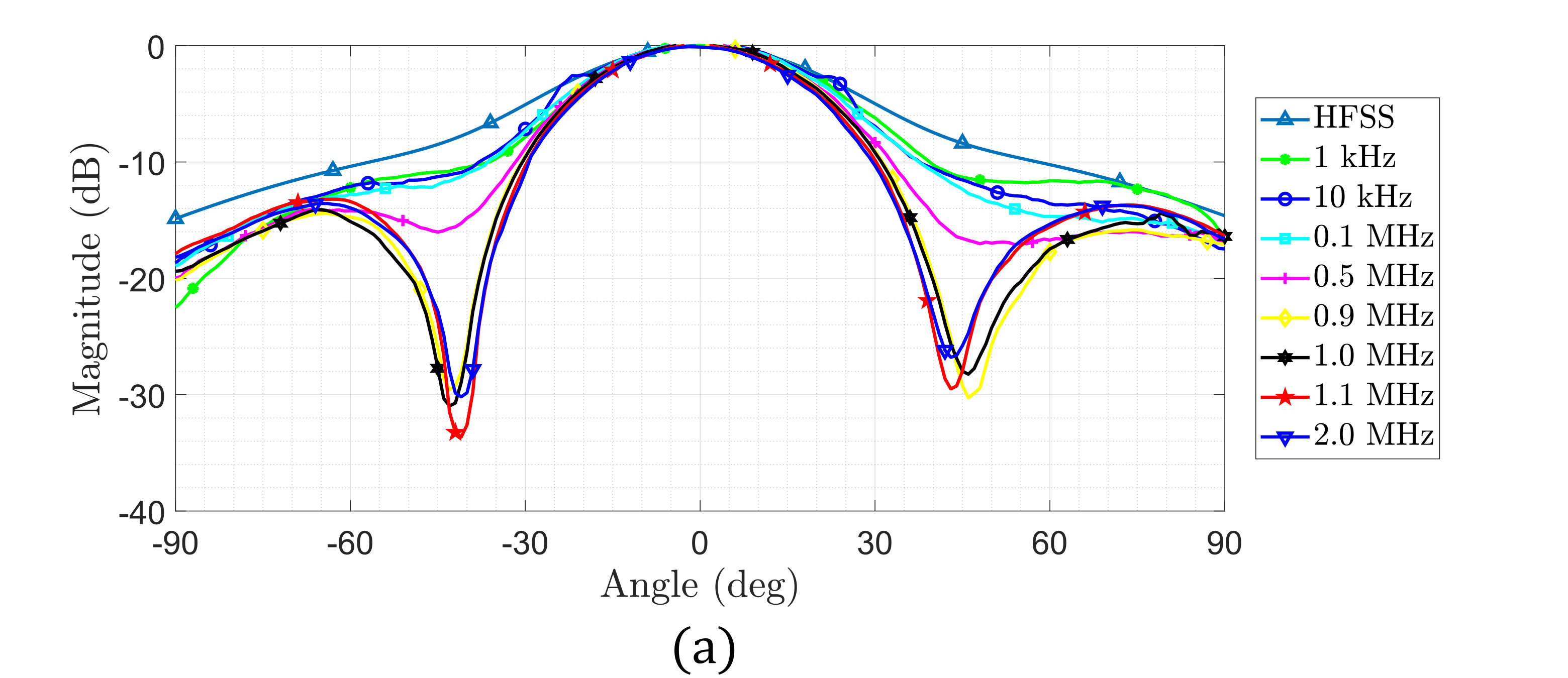}
\includegraphics[width=0.99\columnwidth]{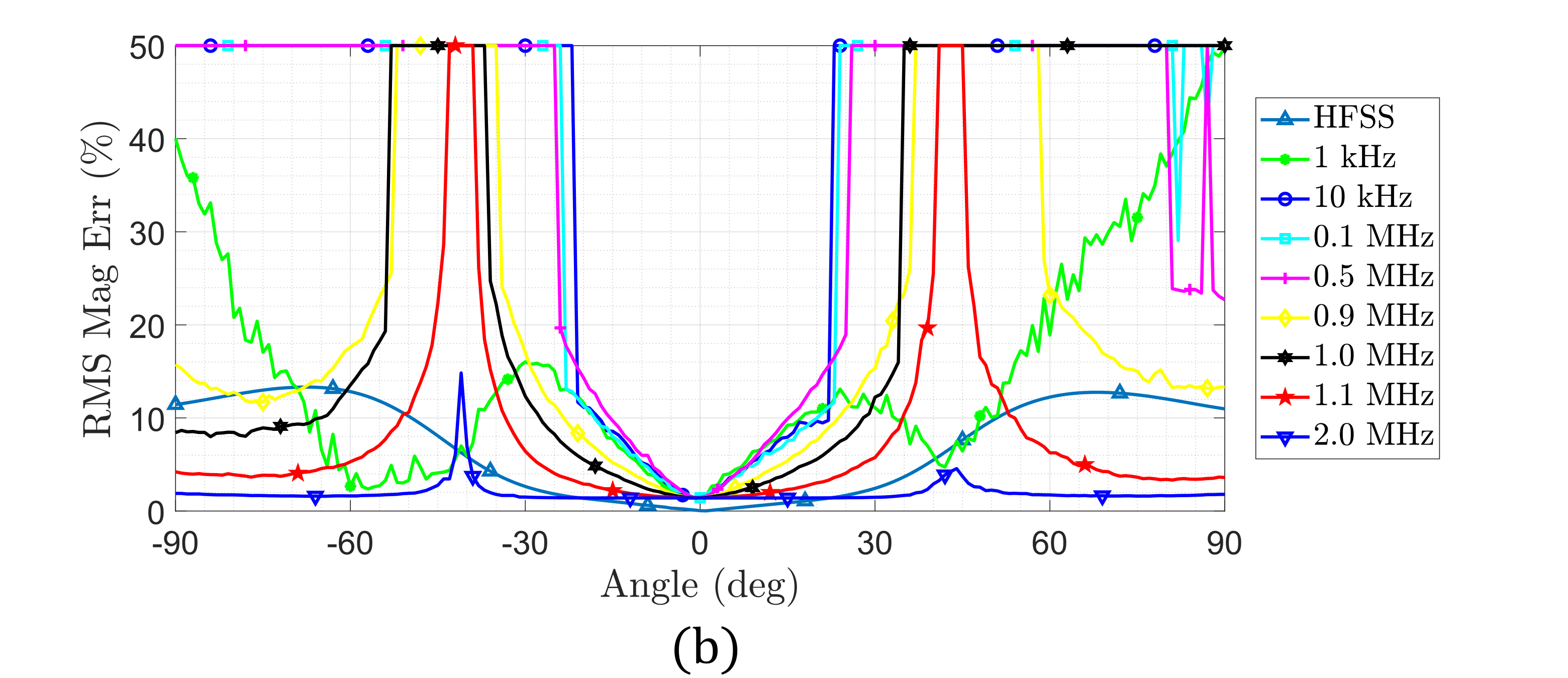}
\includegraphics[width=0.99\columnwidth]{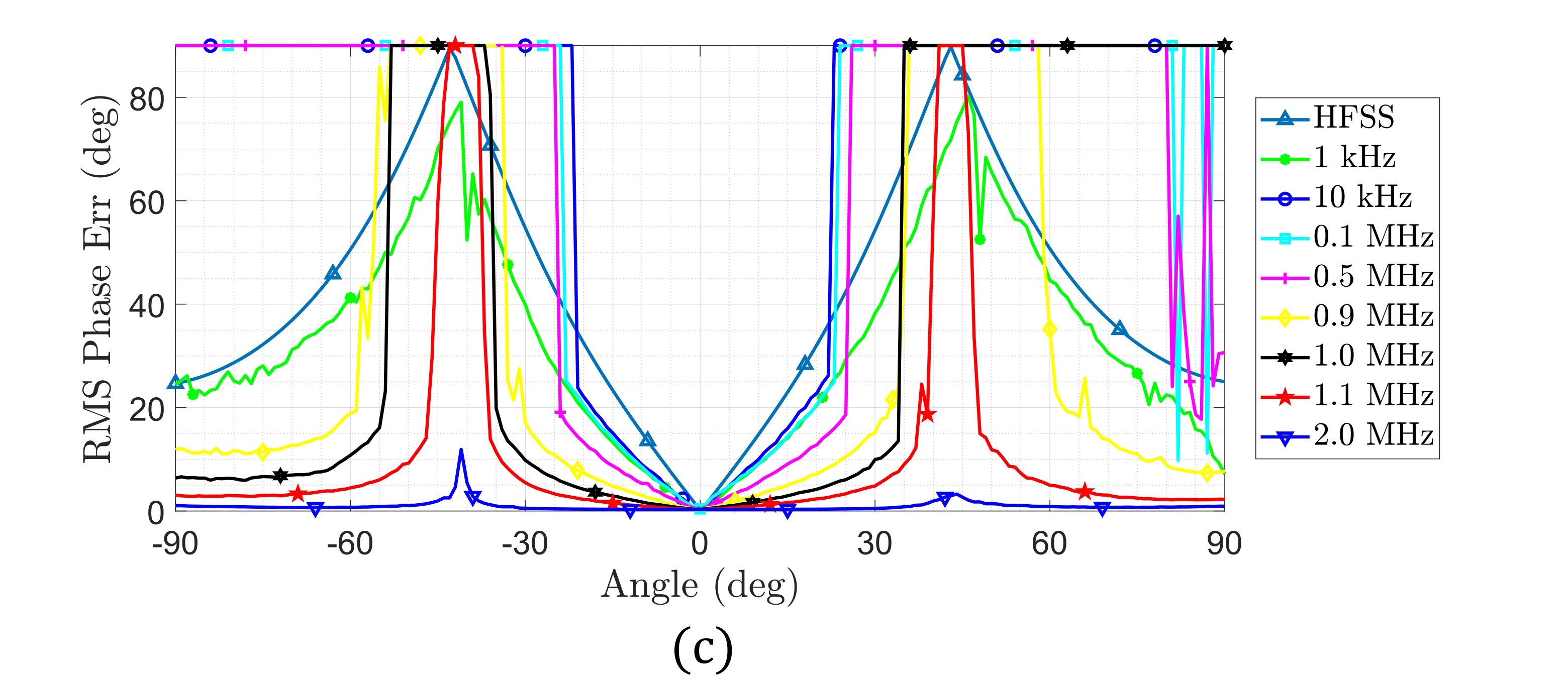}
\includegraphics[width=0.99\columnwidth]{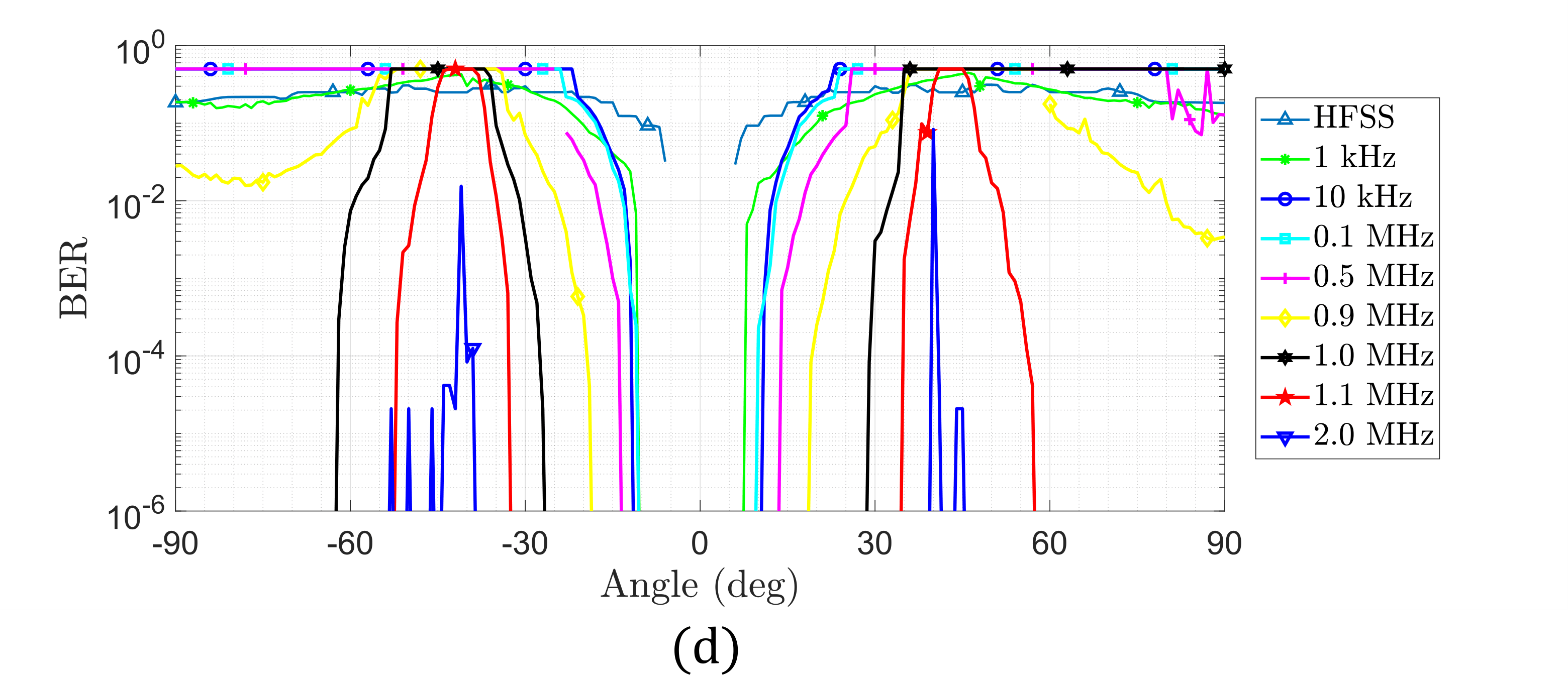}
\caption{RMS power of the baseband signal (a) of the dynamic phased array for various switching frequencies with a 1.5\;MHz receiver bandwidth. The RMS magnitude error $(\%)$ (b) and RMS phase error $(\mathrm{deg})$ (c) between the radiation patterns of the two states show that the regions with large differences (which contribute to wireless security) increase with increasing switching speed. The subsequent bit error ratio (BER) (d) shows the commensurate widening of the information beam as the switching frequency increases. 
}
\label{fig:Dynamic_Antenna_1500kHz_CommsSecurity}
\end{figure}
\begin{figure}[t!]
\centering
\includegraphics[width=0.99\columnwidth]{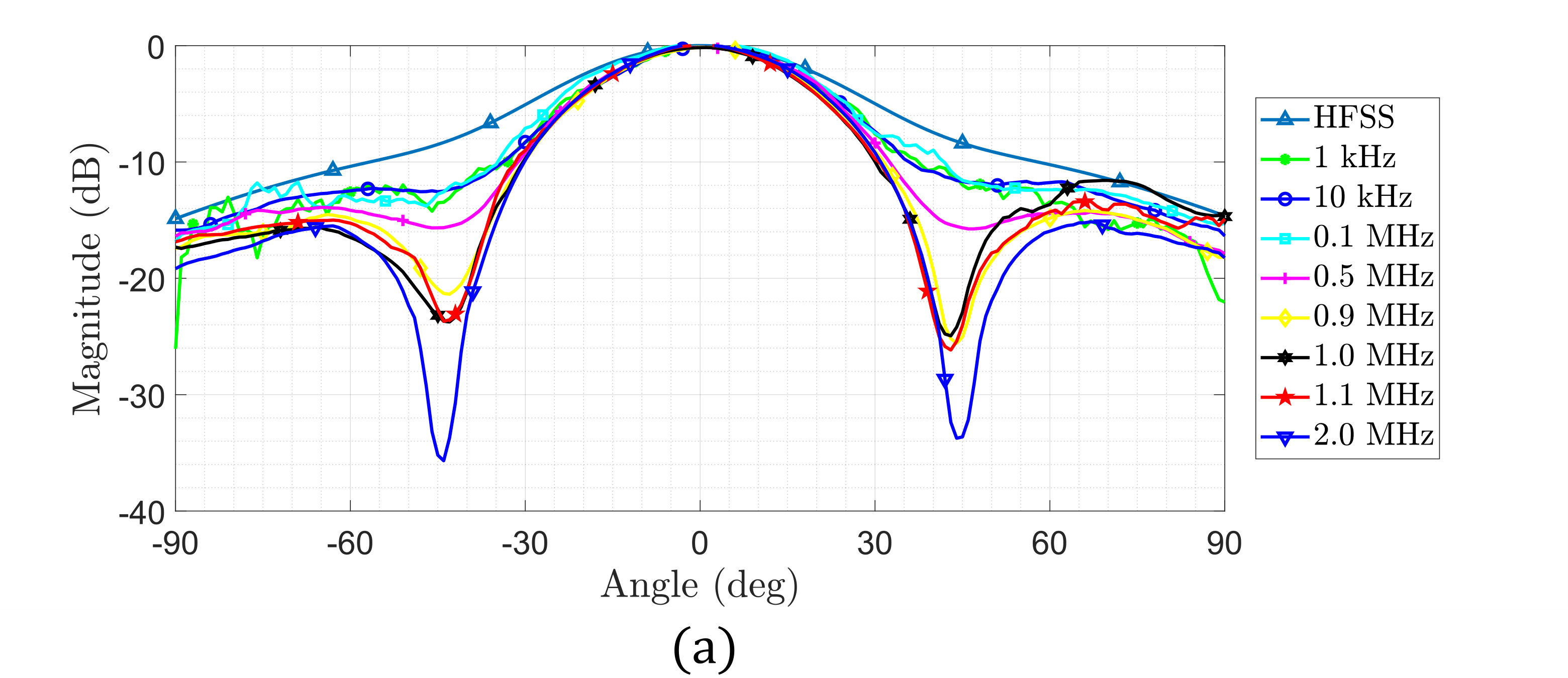}
\includegraphics[width=0.99\columnwidth]{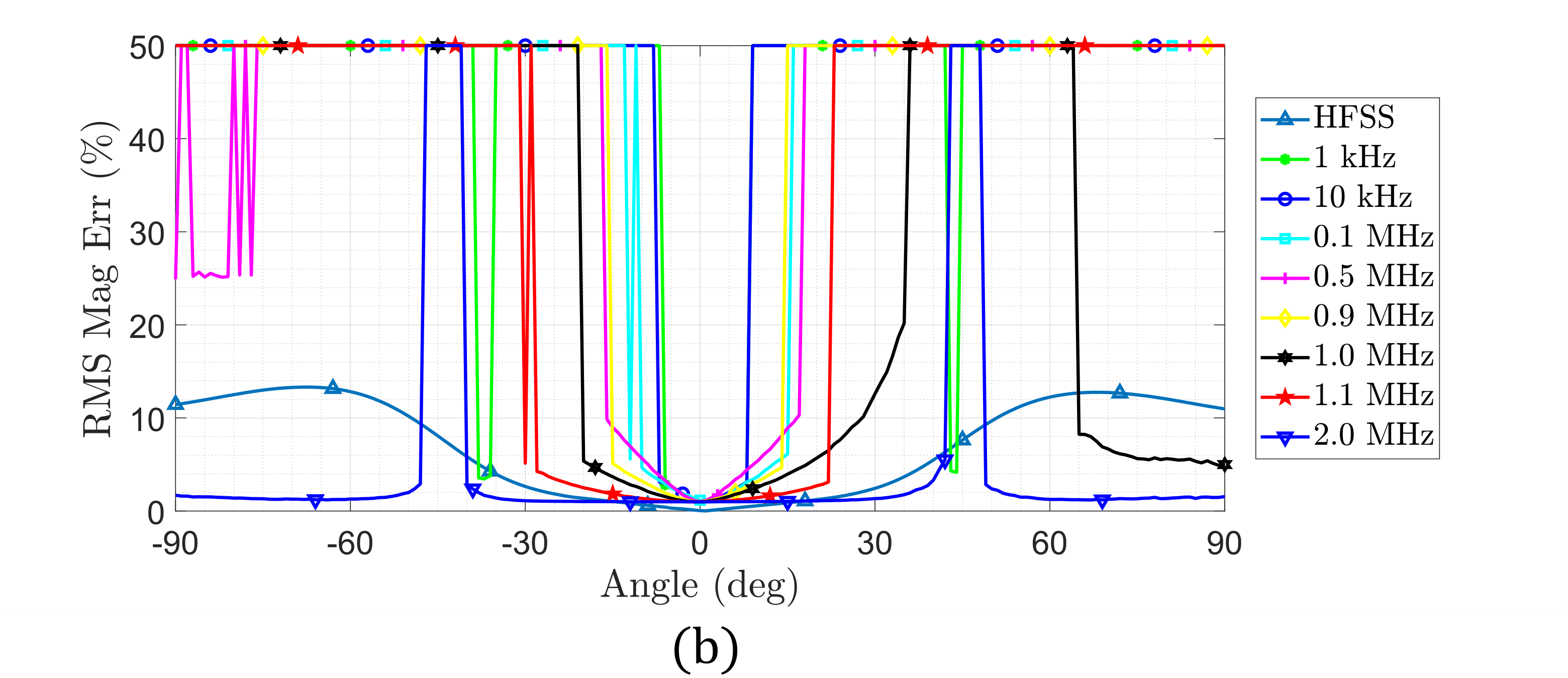}
\includegraphics[width=0.99\columnwidth]{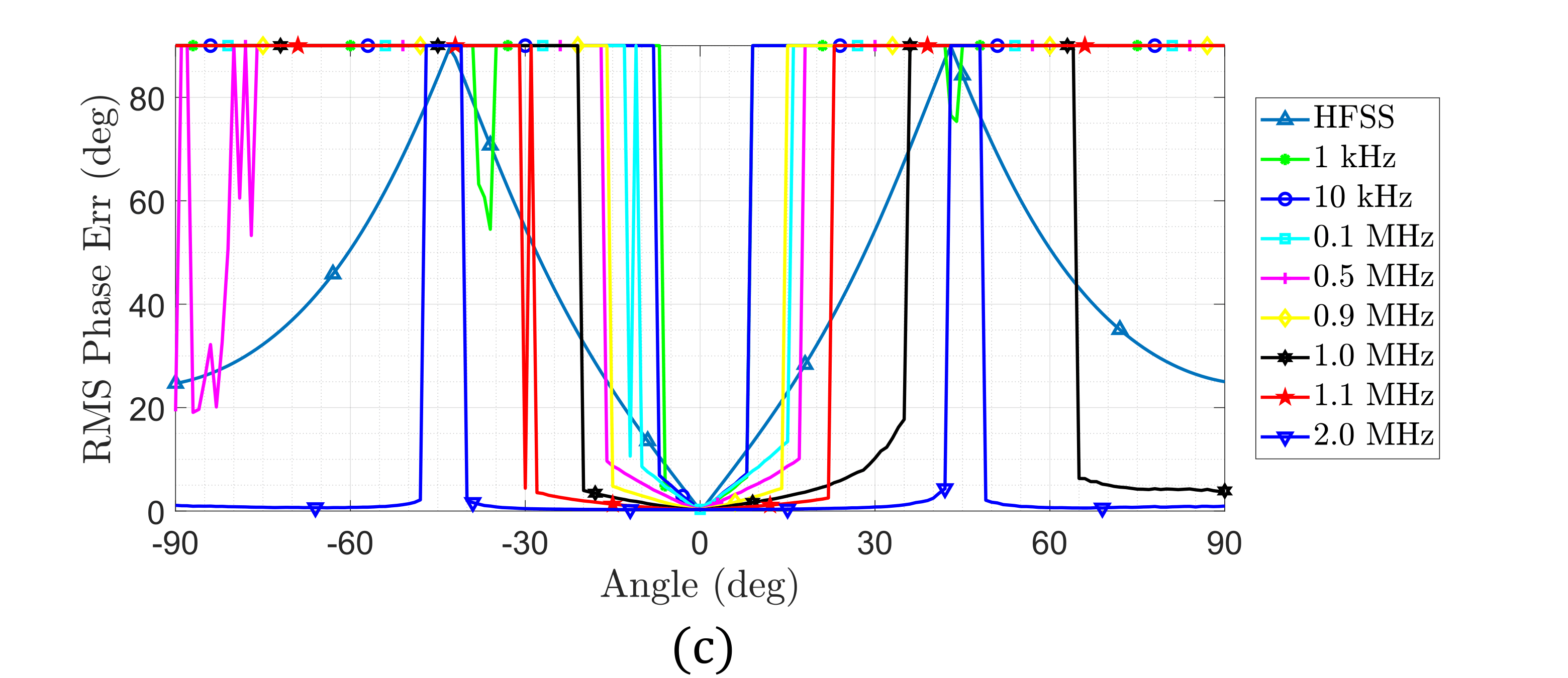}
\includegraphics[width=0.99\columnwidth]{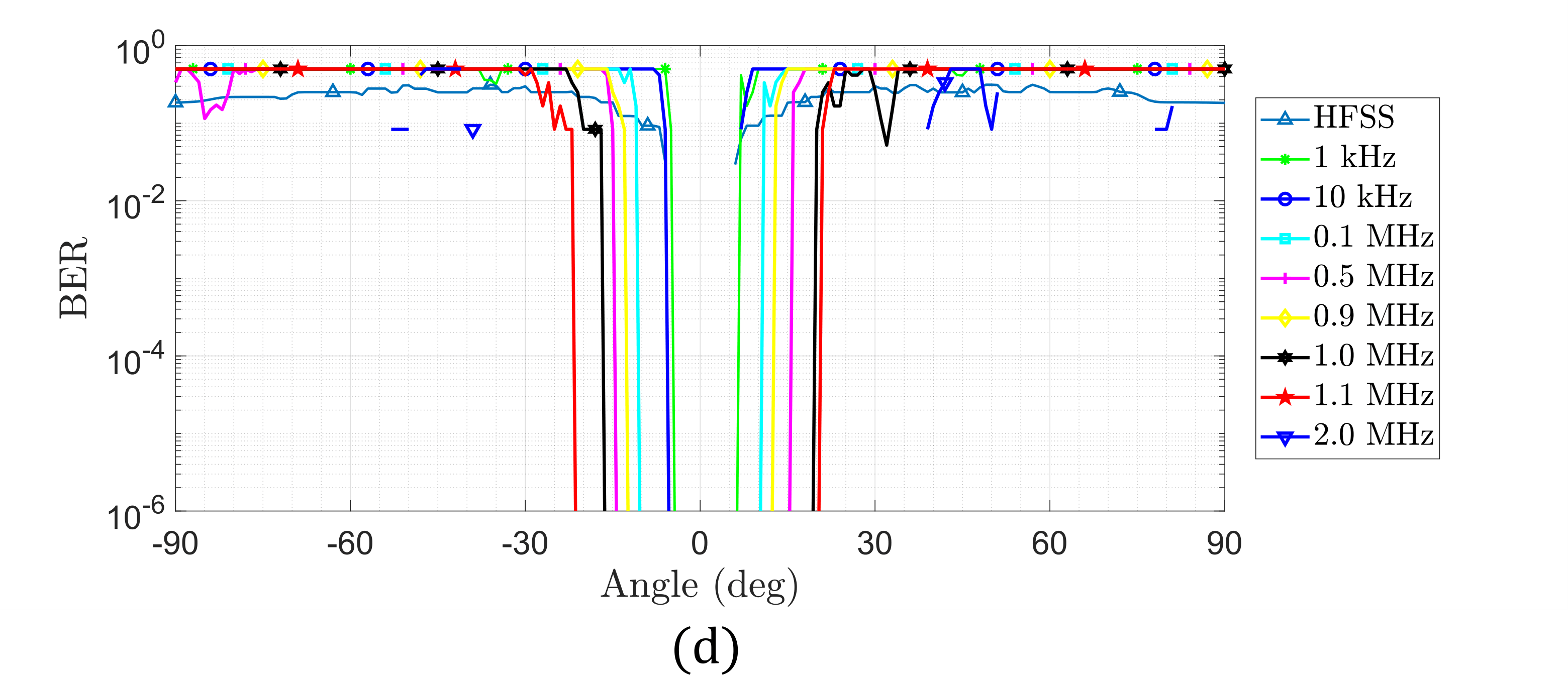}
\caption{RMS power of the baseband signal (a), differential magnitude (b) and phase (c) of the radiation pattern, and subsequent BER (d) for various switching frequencies with a 10\;MHz receiver bandwidth. Compared to Fig.~\ref{fig:Dynamic_Antenna_1500kHz_CommsSecurity}, the information beamwidth is narrower. }
\label{fig:Dynamic_Antenna_8000kHz_CommsSecurity}
\end{figure}

Fig.~\ref{fig:Dynamic_Antenna_Pattern_SimvsMeas} shows a comparison of the theoretical model and the heuristic model as a function of beamsteering angle. Specifically, the absolute error as a function of angle and steering angle is shown for the fundamental component~\eqref{eqn:TimeAveraged_ArrayFactor} and the sideband signals~\eqref{eqn:Harmonic_Power_Theory1}. The points of maximum error between the models corresponds to the nulls of the fundamental term and the nulls of the sideband terms. The sideband terms have more error than the time-averaged response, which can be attributed to the discrete derivative being less accurate; however, the RMSE across all angles was less than 0.3\% for either equation and the peak error was less than 2\% (concentrated at the null; the error is significantly lower elsewhere), validating the accuracy of the heuristic model.
Measurements of the fundamental and sideband components were conducted at three different steering angle and compared to the two models in Fig.~\ref{fig:Dynamic_Antenna_Array_Model_SimvsMeasured}. 
Taking the average PSD over the signal bandwidth of both the time-averaged response and the intermodulation product, we can see the time-averaged power over angle $\phi$ in Fig.~\ref{fig:Dynamic_Antenna_Array_Model_SimvsMeasured} (a), (c), (e), and sideband signal power in Fig. \ref{fig:Dynamic_Antenna_Array_Model_SimvsMeasured} (b), (d), (f). The measured power follows the simulated response well; however, there is more error near endfire, which could be attributed to environmental scattering, mutual coupling between the antennas near endfire, or minor fabrication differences between the two patch antennas.



\section{Analysis of Switching Rate, Receiver Bandwidth, and Information Beamsteering}

\begin{figure*}
\includegraphics[width=0.237\textwidth]{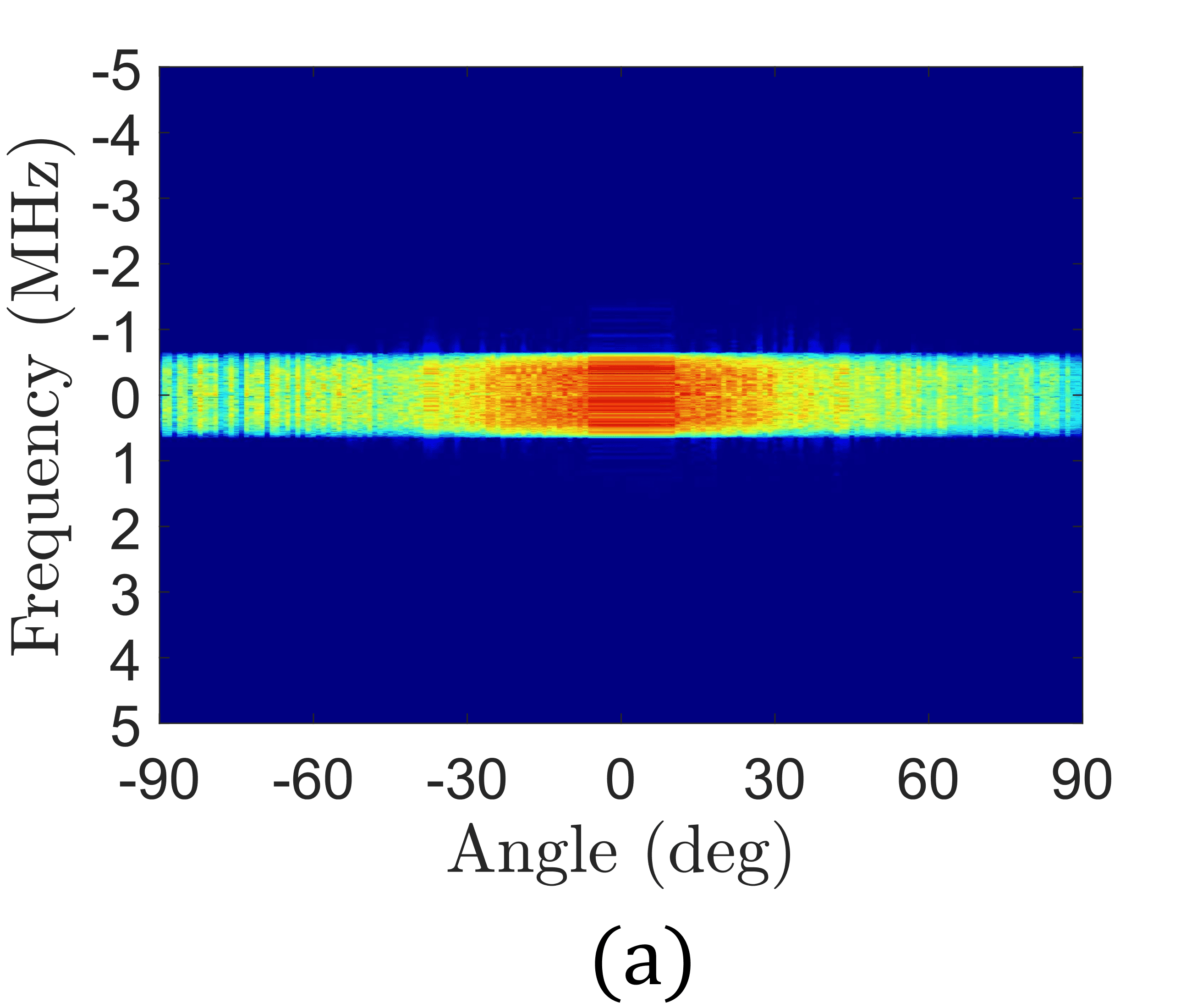}
\includegraphics[width=0.237\textwidth]{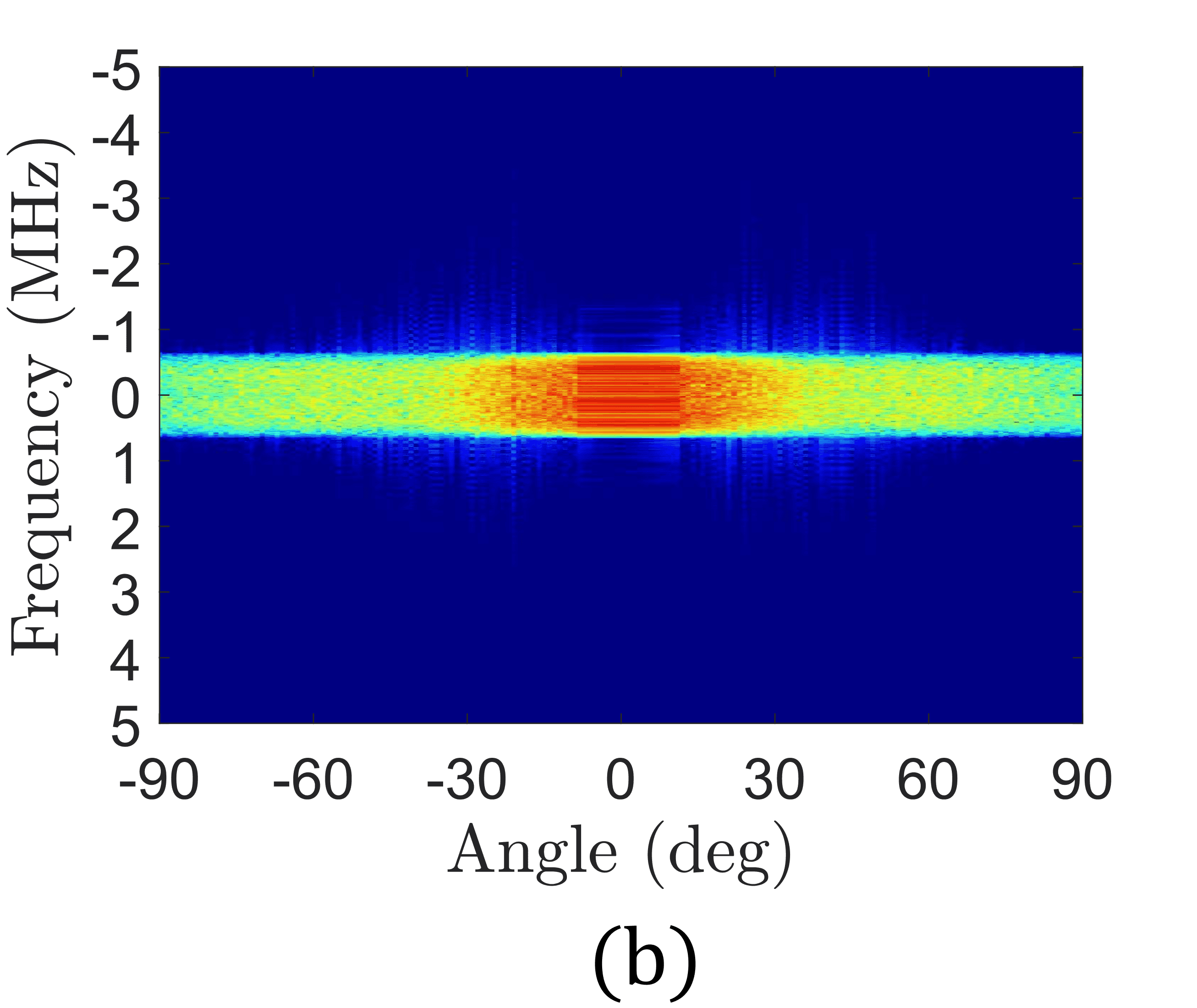}
\includegraphics[width=0.237\textwidth]{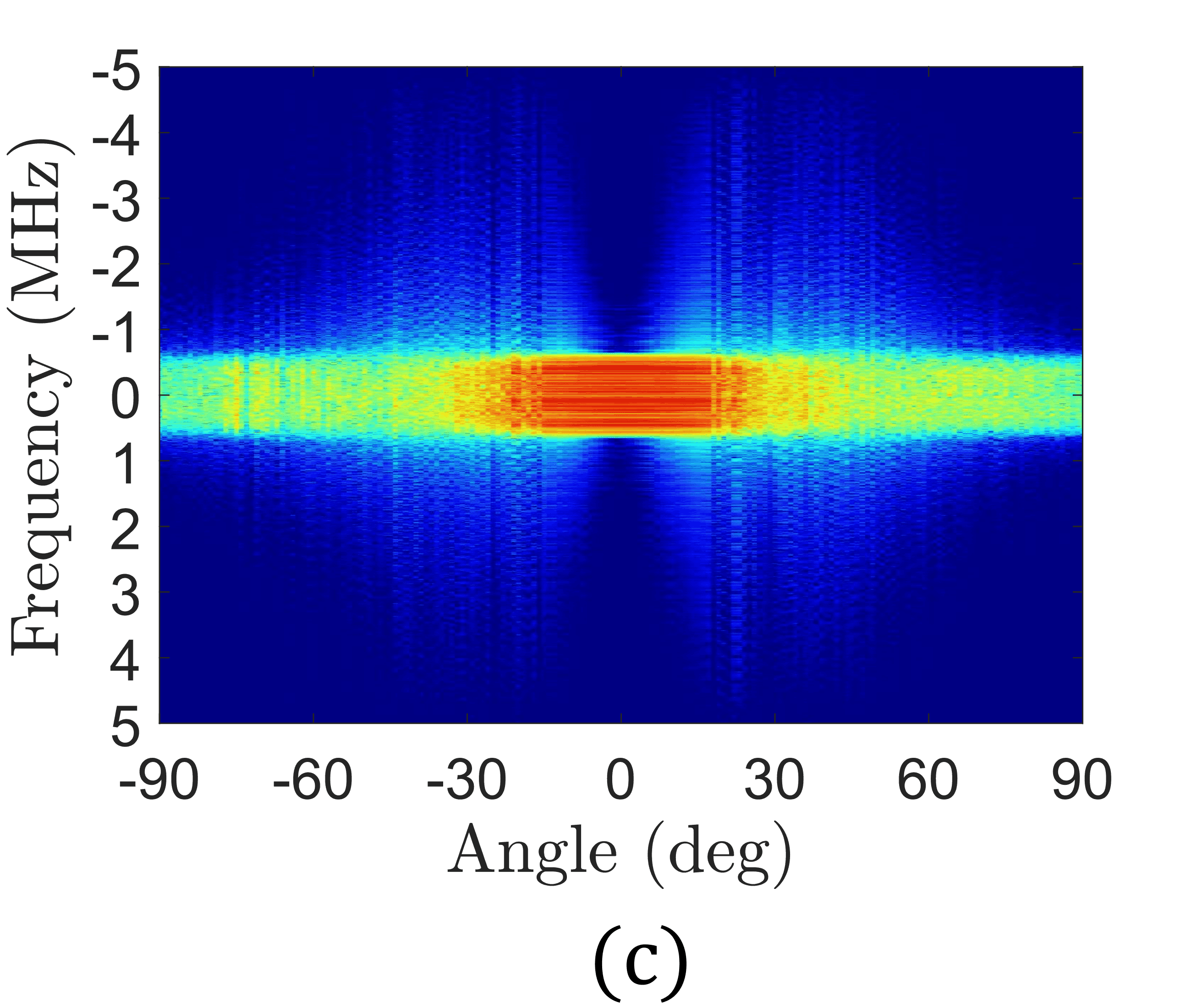}
\includegraphics[width=0.237\textwidth]{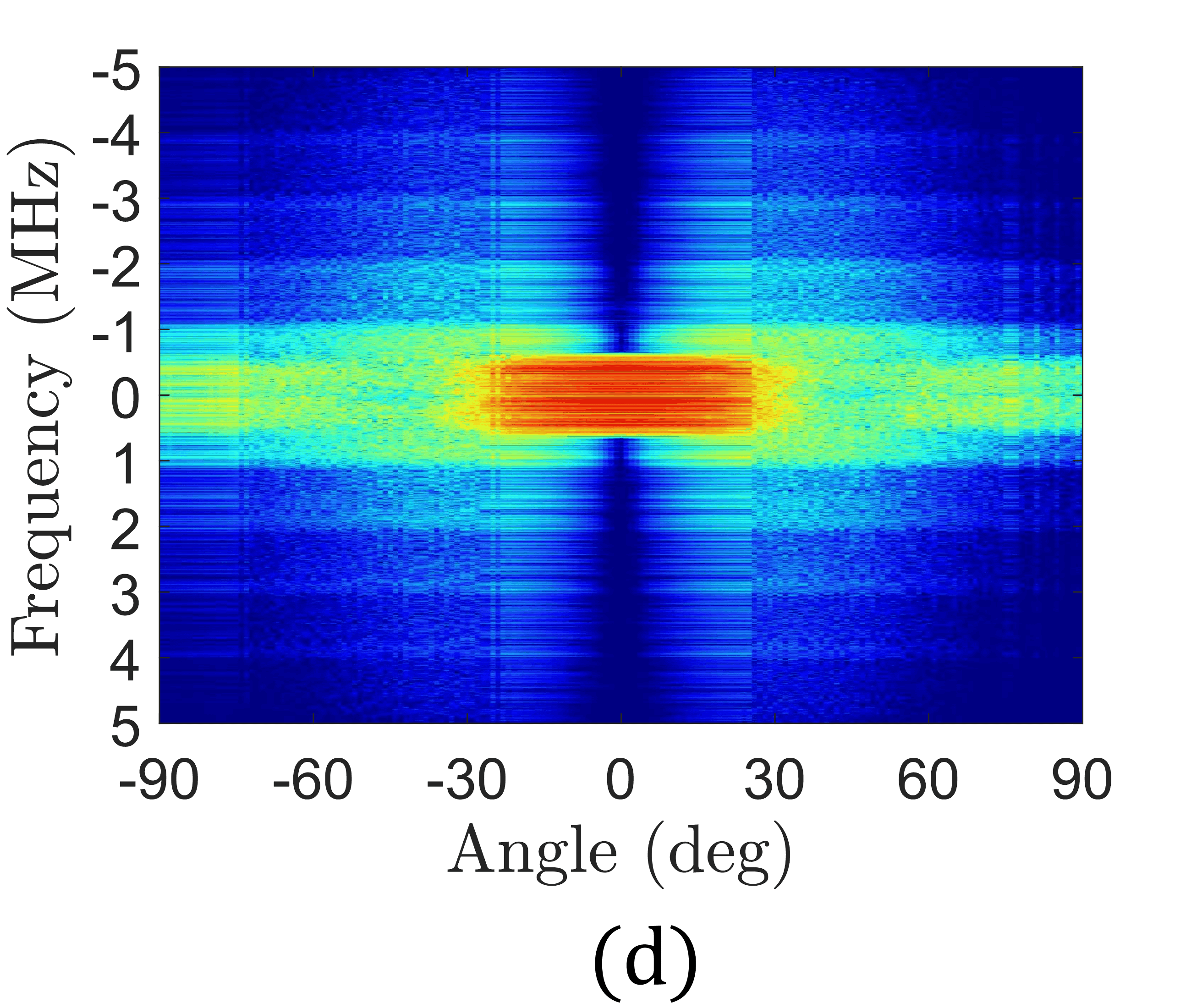}
\includegraphics[width=0.0324\textwidth]{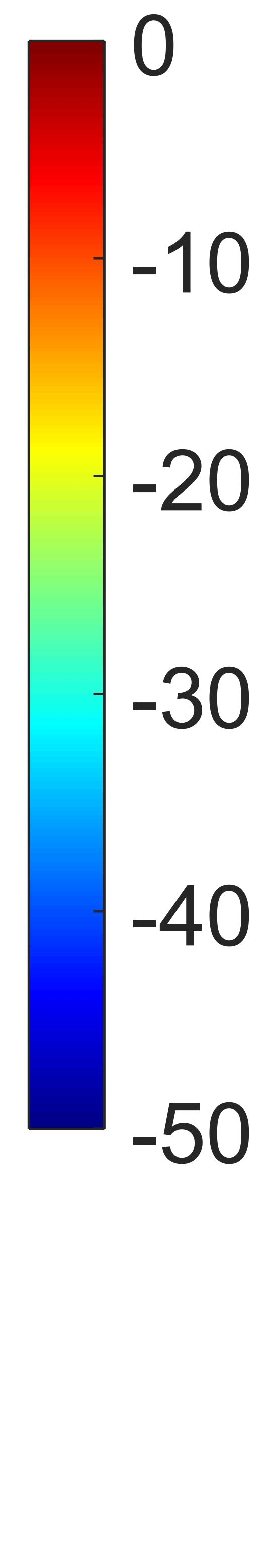}
\includegraphics[width=0.237\textwidth]{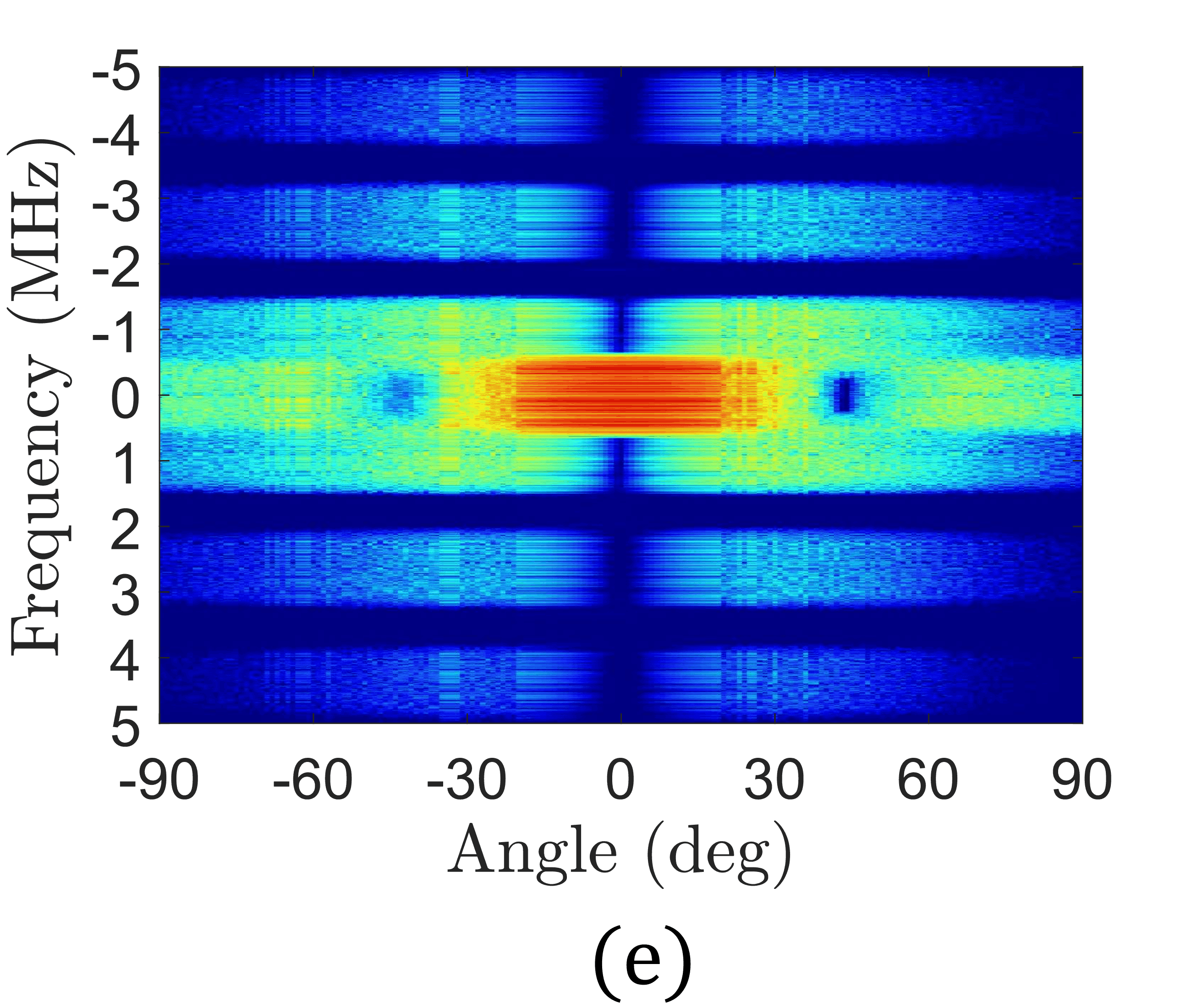}
\includegraphics[width=0.237\textwidth]{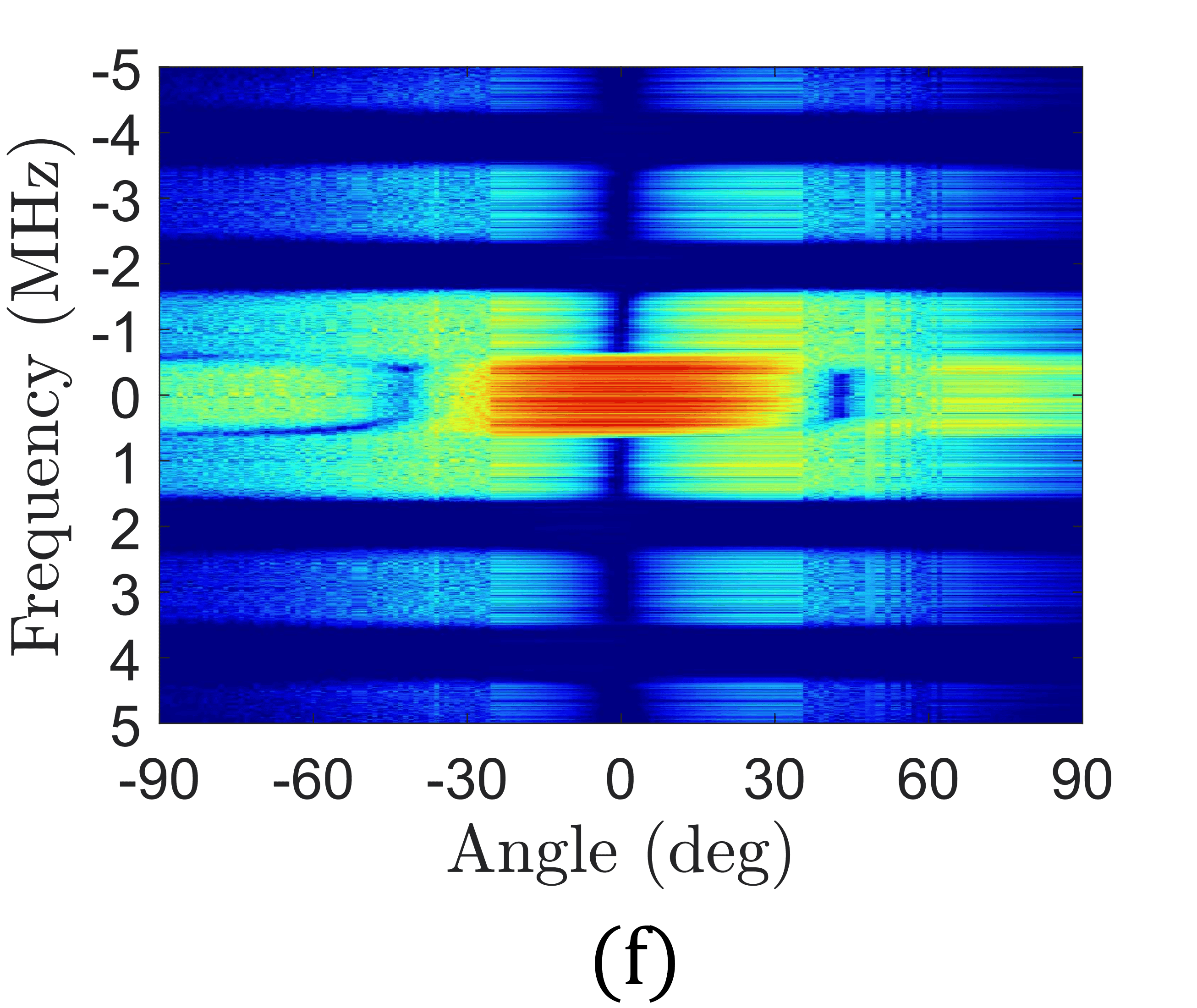}
\includegraphics[width=0.237\textwidth]{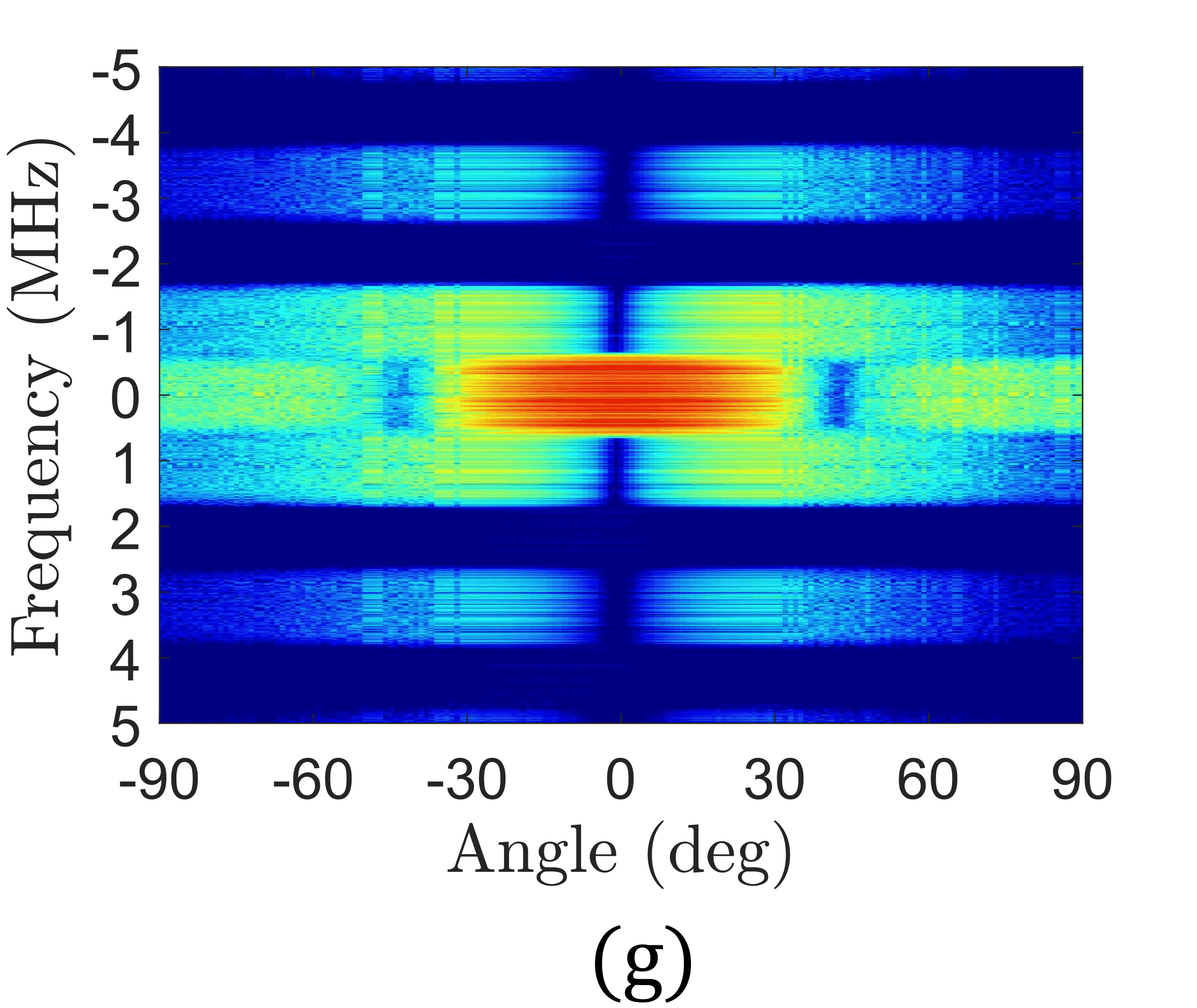}
\includegraphics[width=0.237\textwidth]{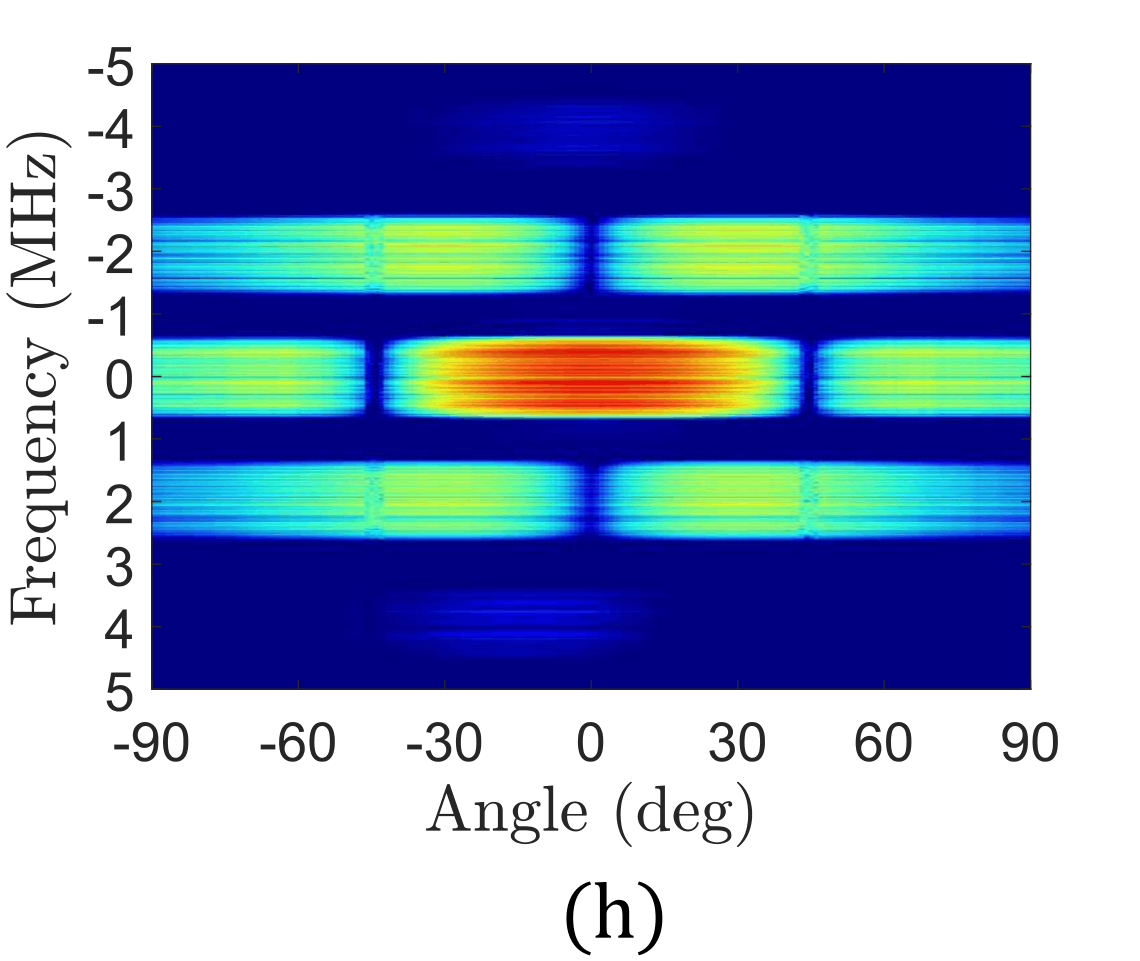}
\includegraphics[width=0.0324\textwidth]{Colorbar-1}
\caption{Normalized power spectral density (dB) of the dynamic two-element phased array switching at a rate of (top left to bottom right) 1\;kHz, 10\;kHz, 100\;kHz, 500\;kHz, 900\;kHz, 1.0\;MHz, 1.1\;MHz, and 2.0\;MHz. The wider bandwidth of 10\;MHz leads to capturing more spectral sideband information, leading to narrower information beamwidths.
}
\label{fig:PSD_8000MHz}
\end{figure*}

\begin{figure*}
\centering
\includegraphics[width=0.99\textwidth]{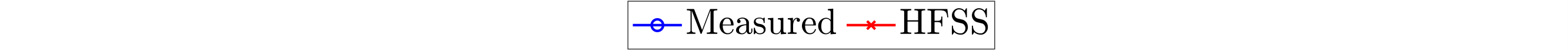}
\includegraphics[width=0.195\textwidth]{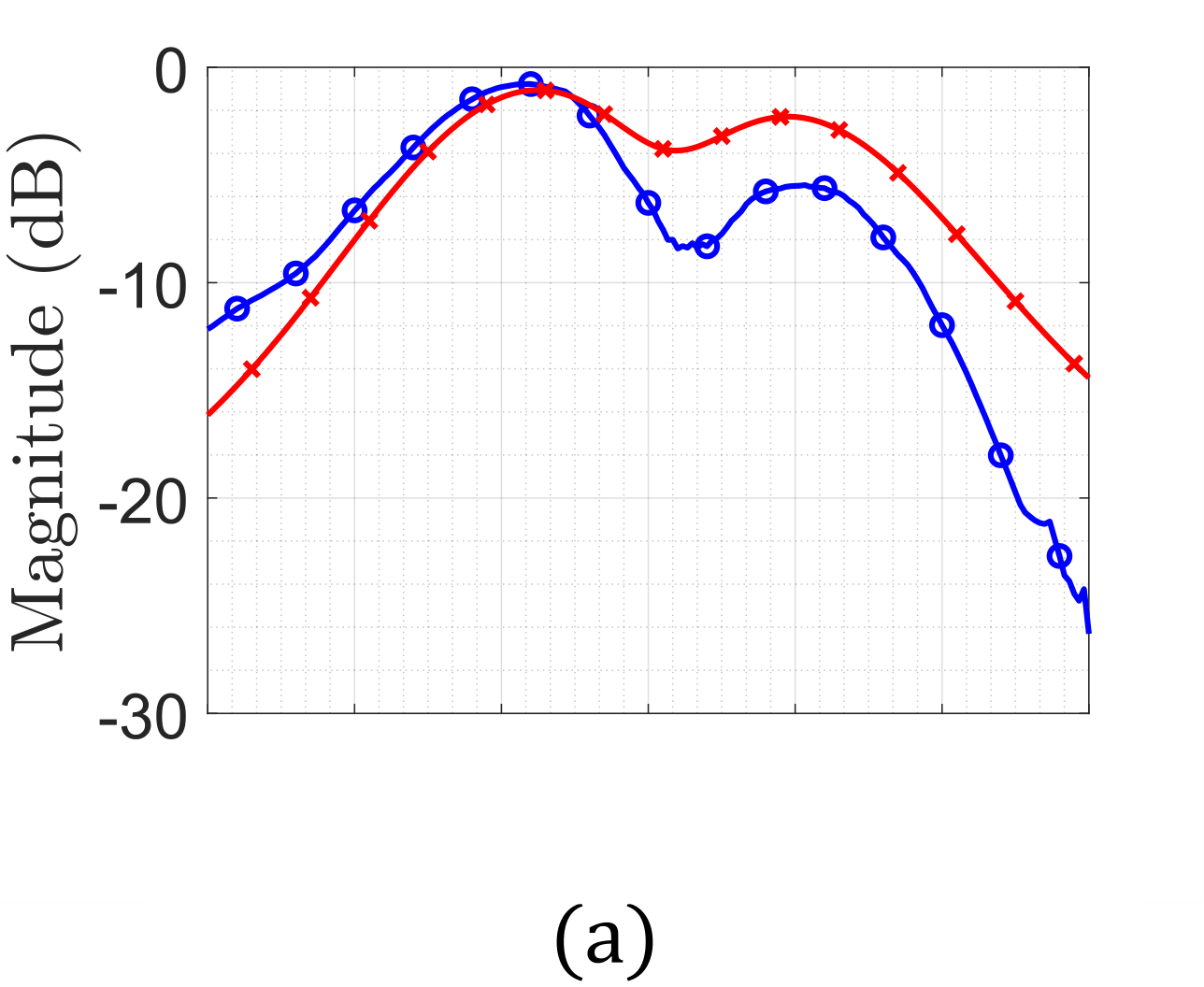}
\includegraphics[width=0.195\textwidth]{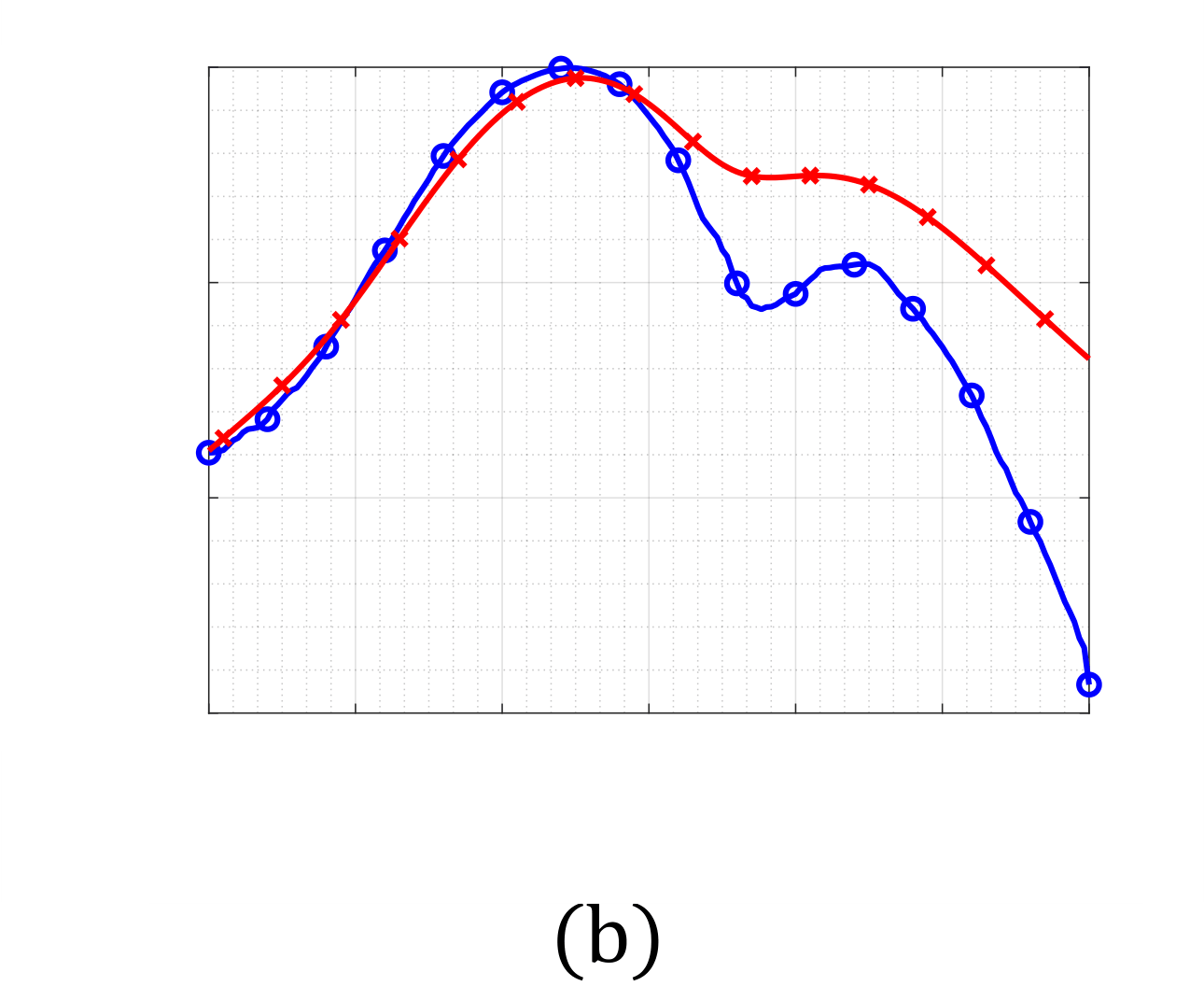}
\includegraphics[width=0.195\textwidth]{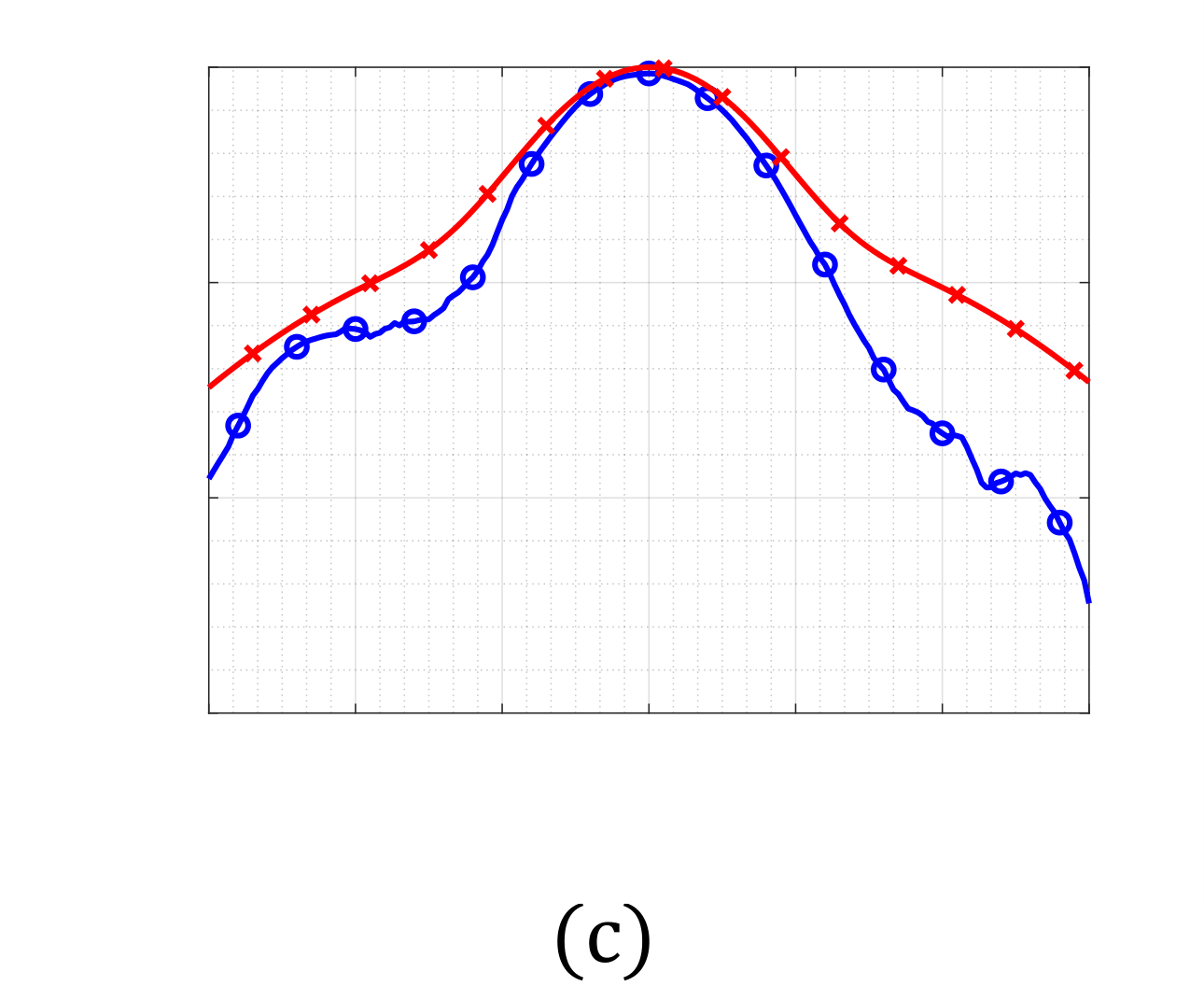}
\includegraphics[width=0.195\textwidth]{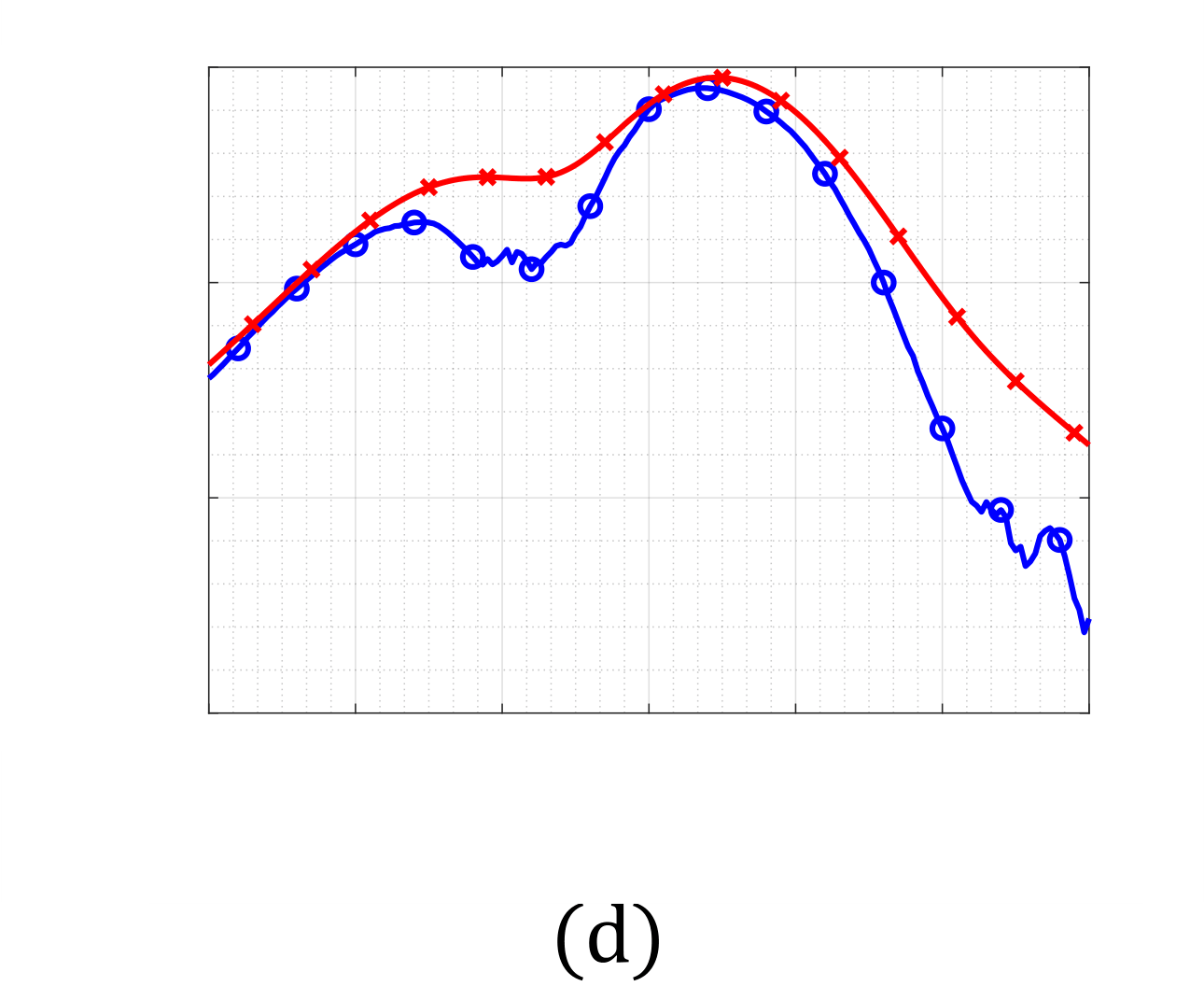}
\includegraphics[width=0.195\textwidth]{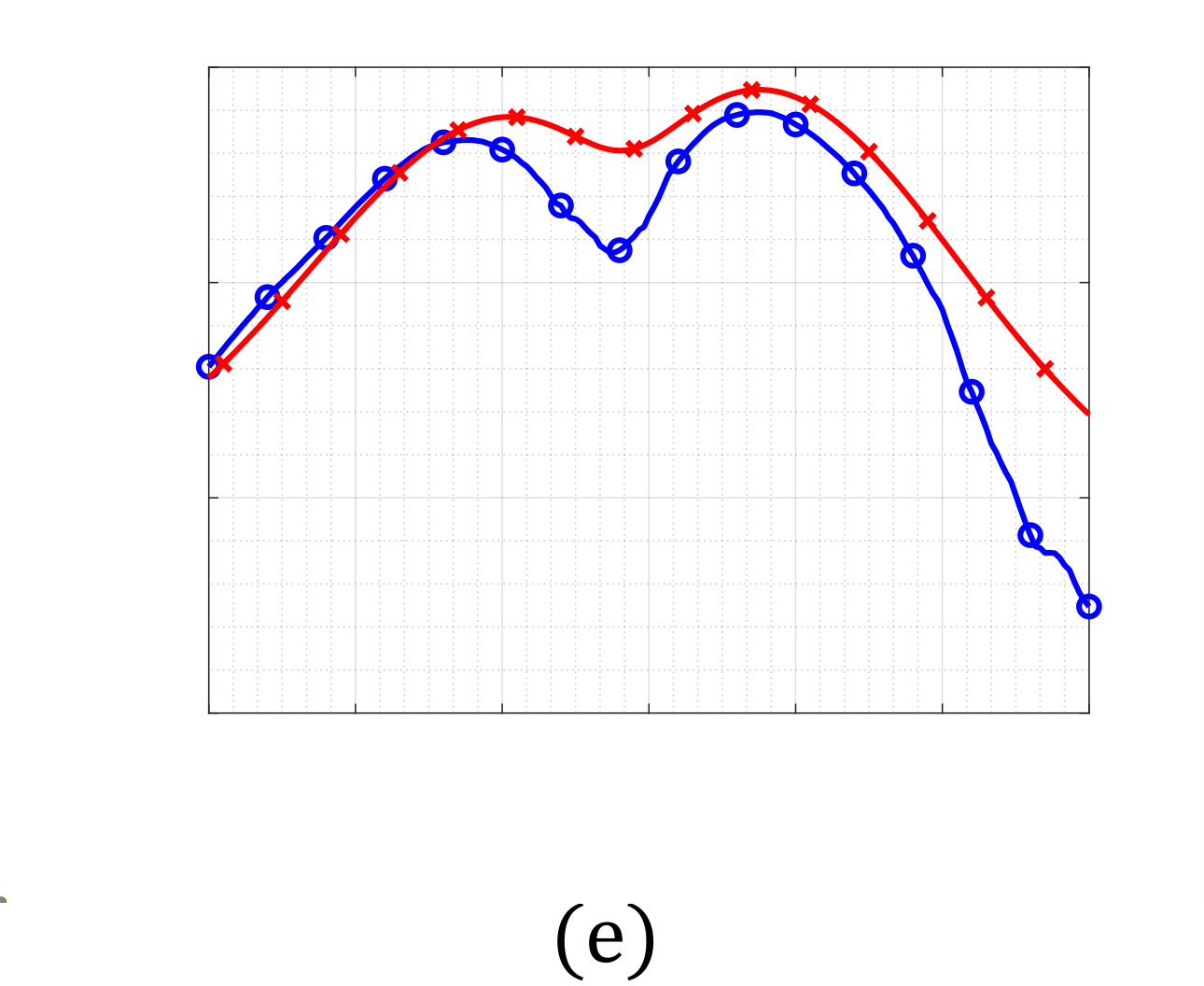}

\includegraphics[width=0.195\textwidth]{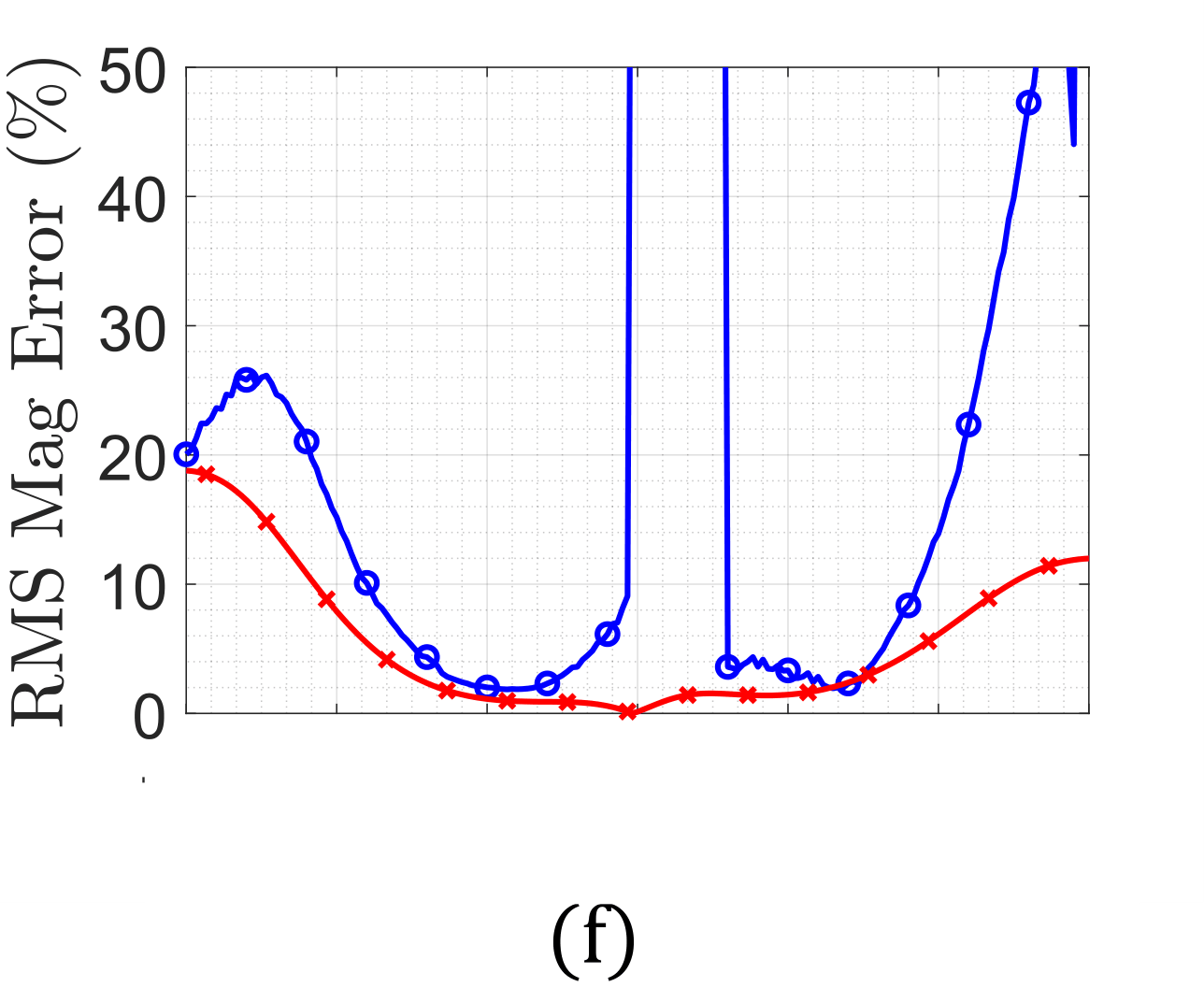}
\includegraphics[width=0.195\textwidth]{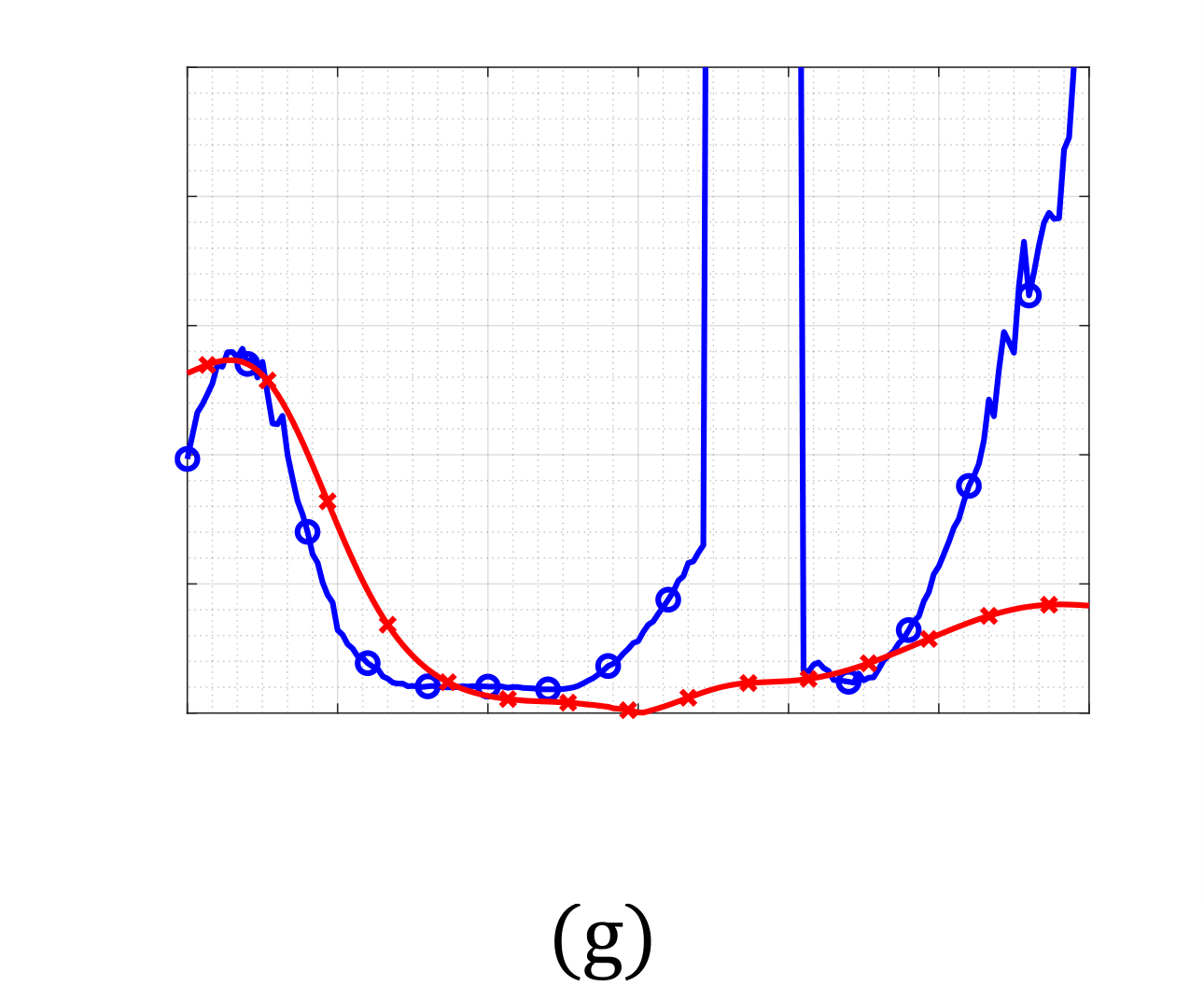}
\includegraphics[width=0.195\textwidth]{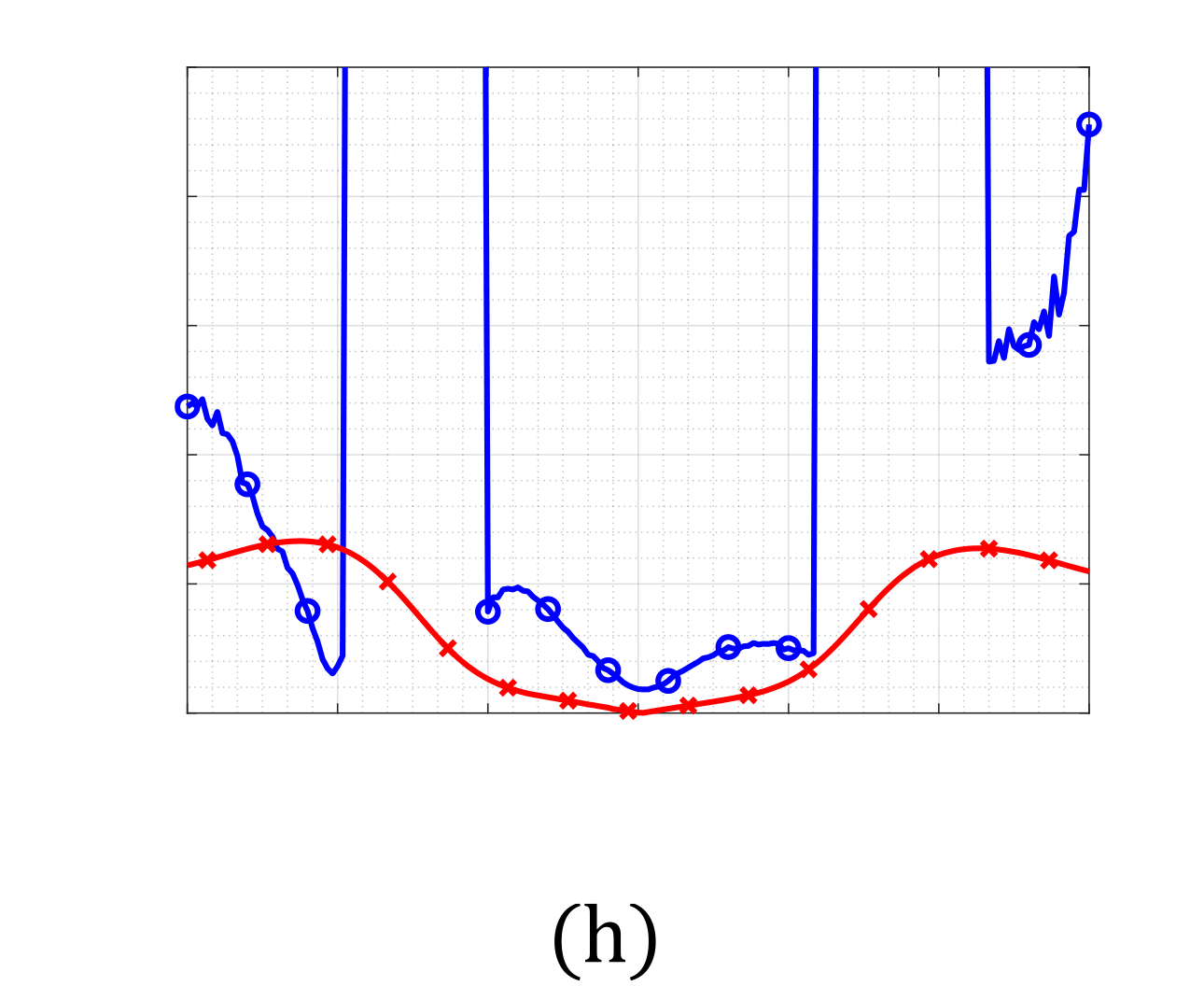}
\includegraphics[width=0.195\textwidth]{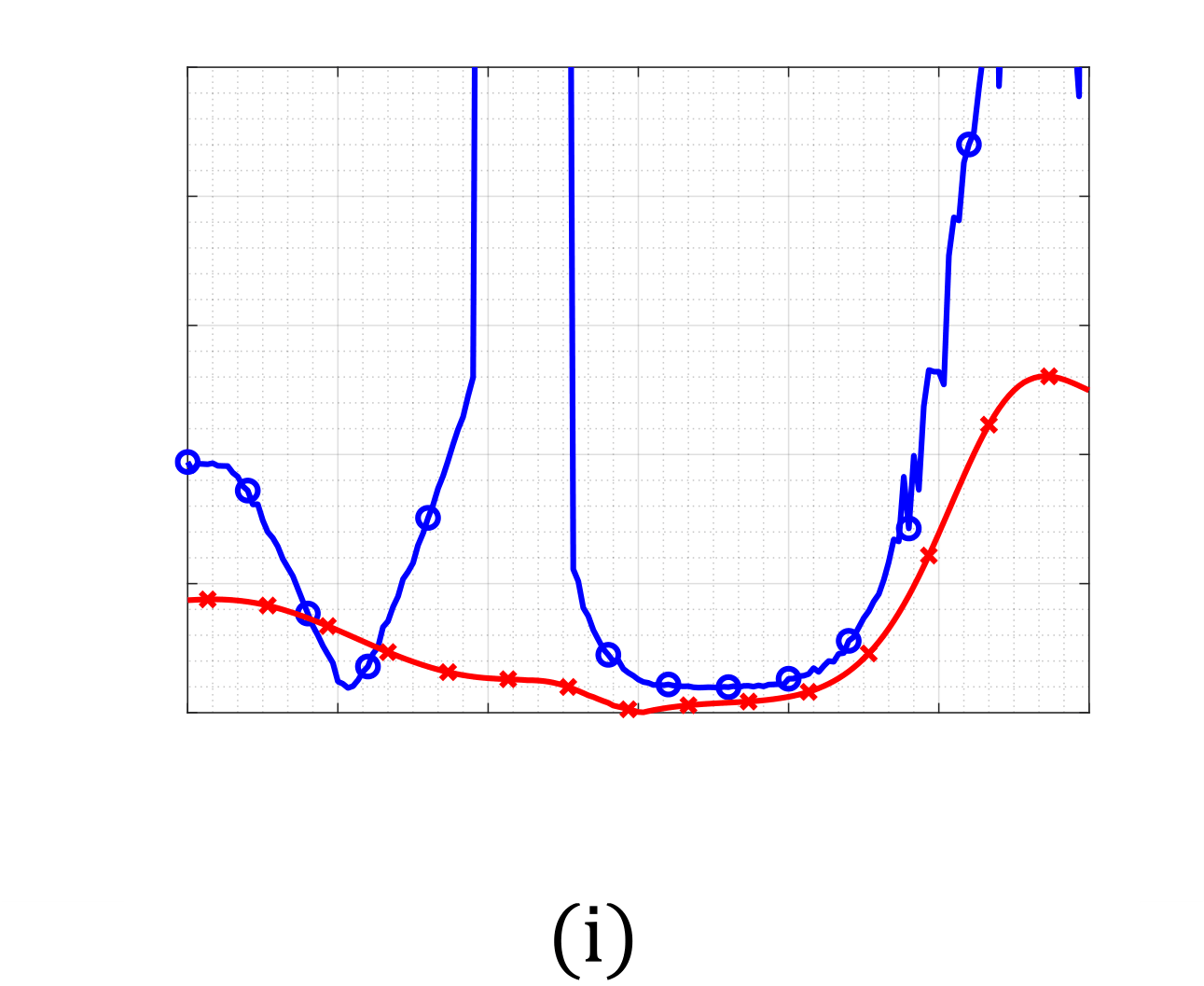}
\includegraphics[width=0.195\textwidth]{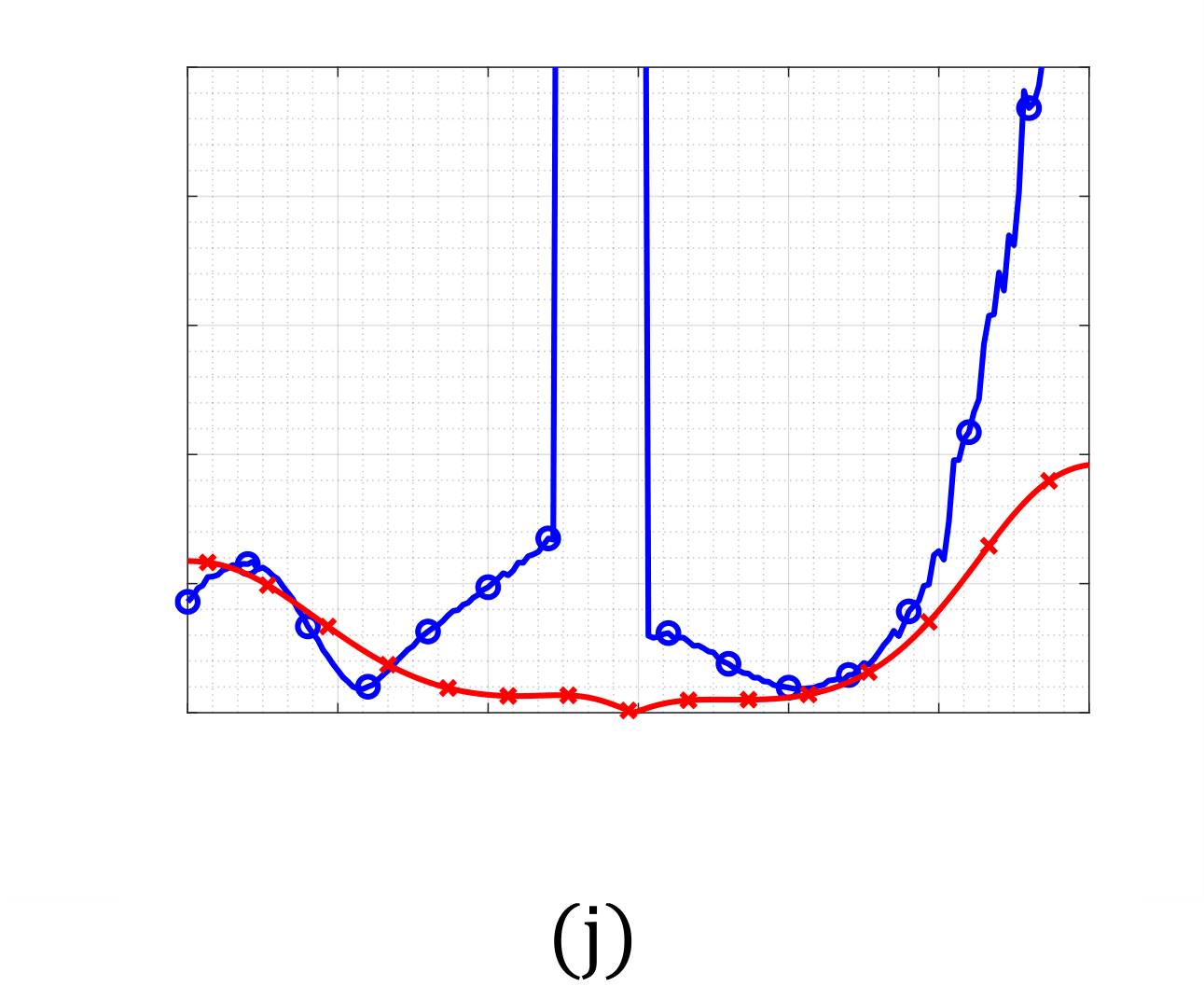}

\includegraphics[width=0.195\textwidth]{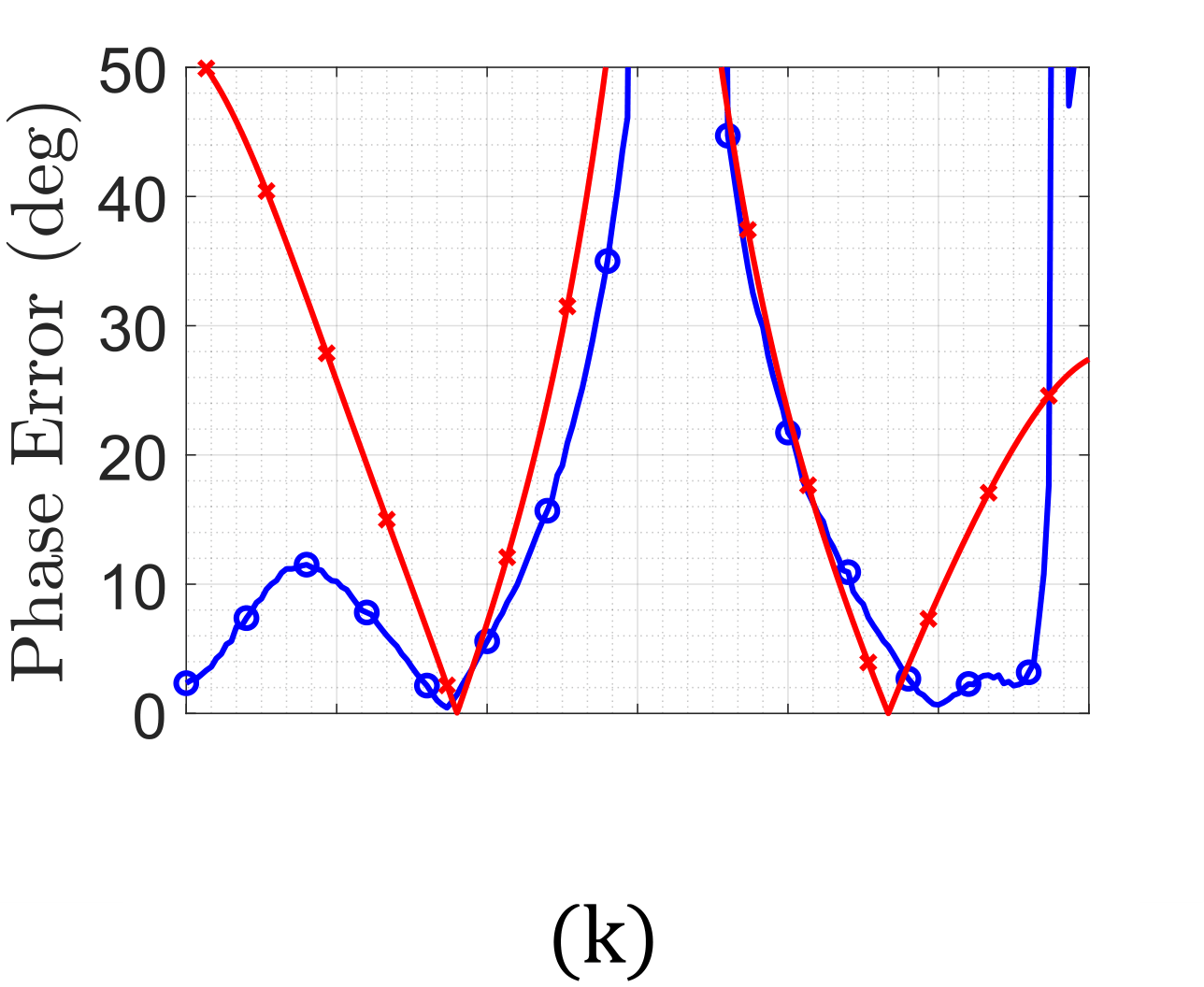}
\includegraphics[width=0.195\textwidth]{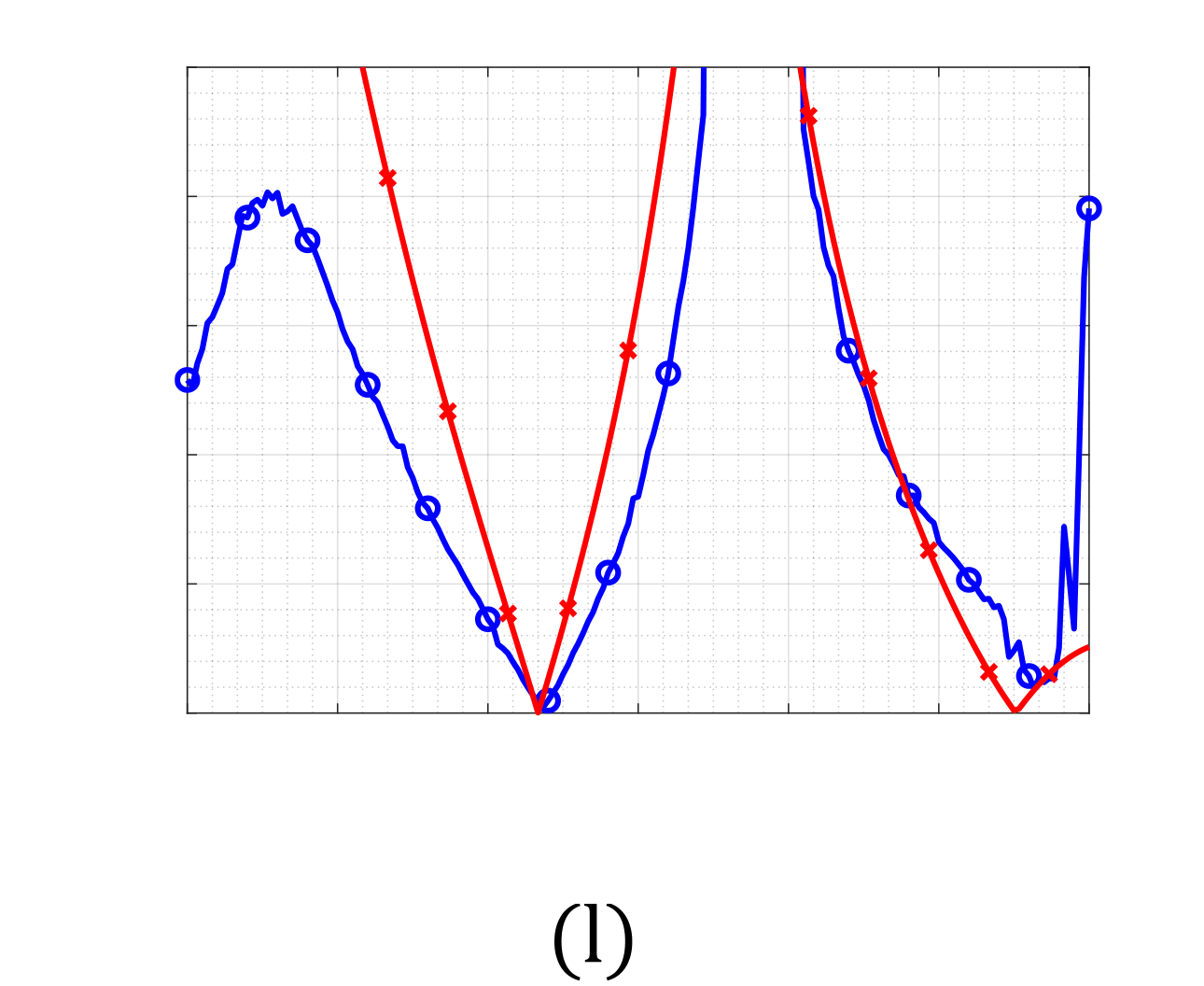}
\includegraphics[width=0.195\textwidth]{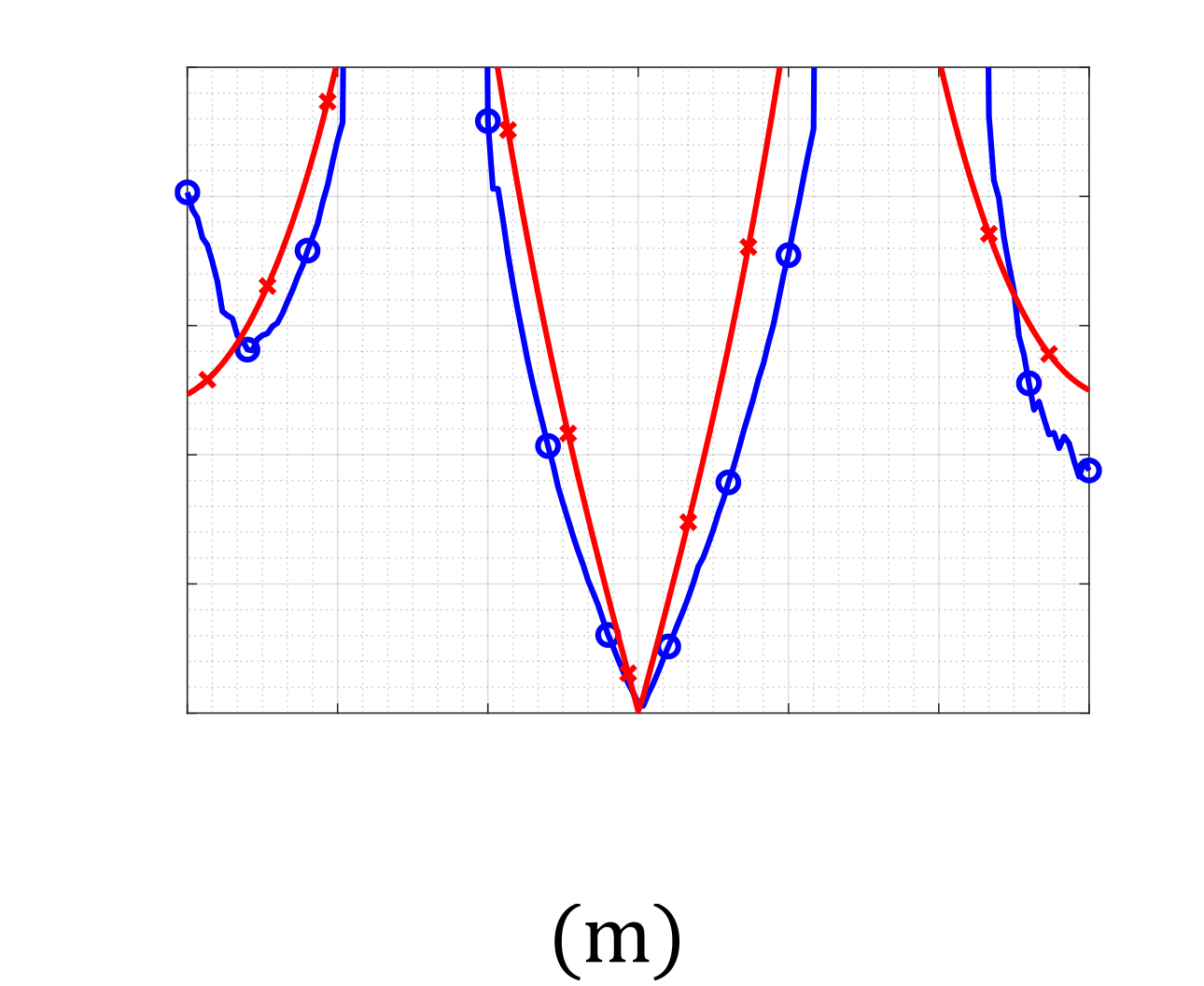}
\includegraphics[width=0.195\textwidth]{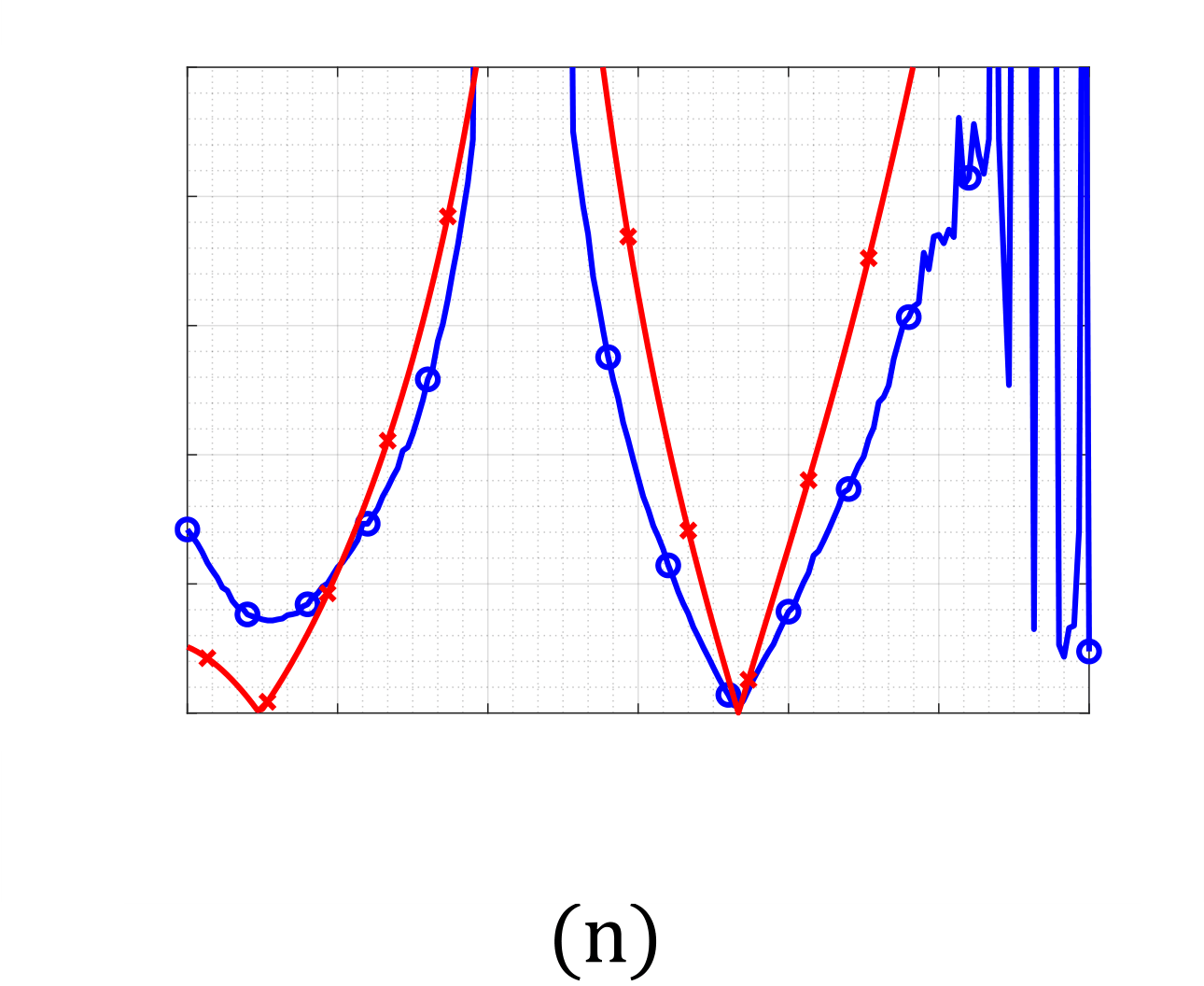}
\includegraphics[width=0.195\textwidth]{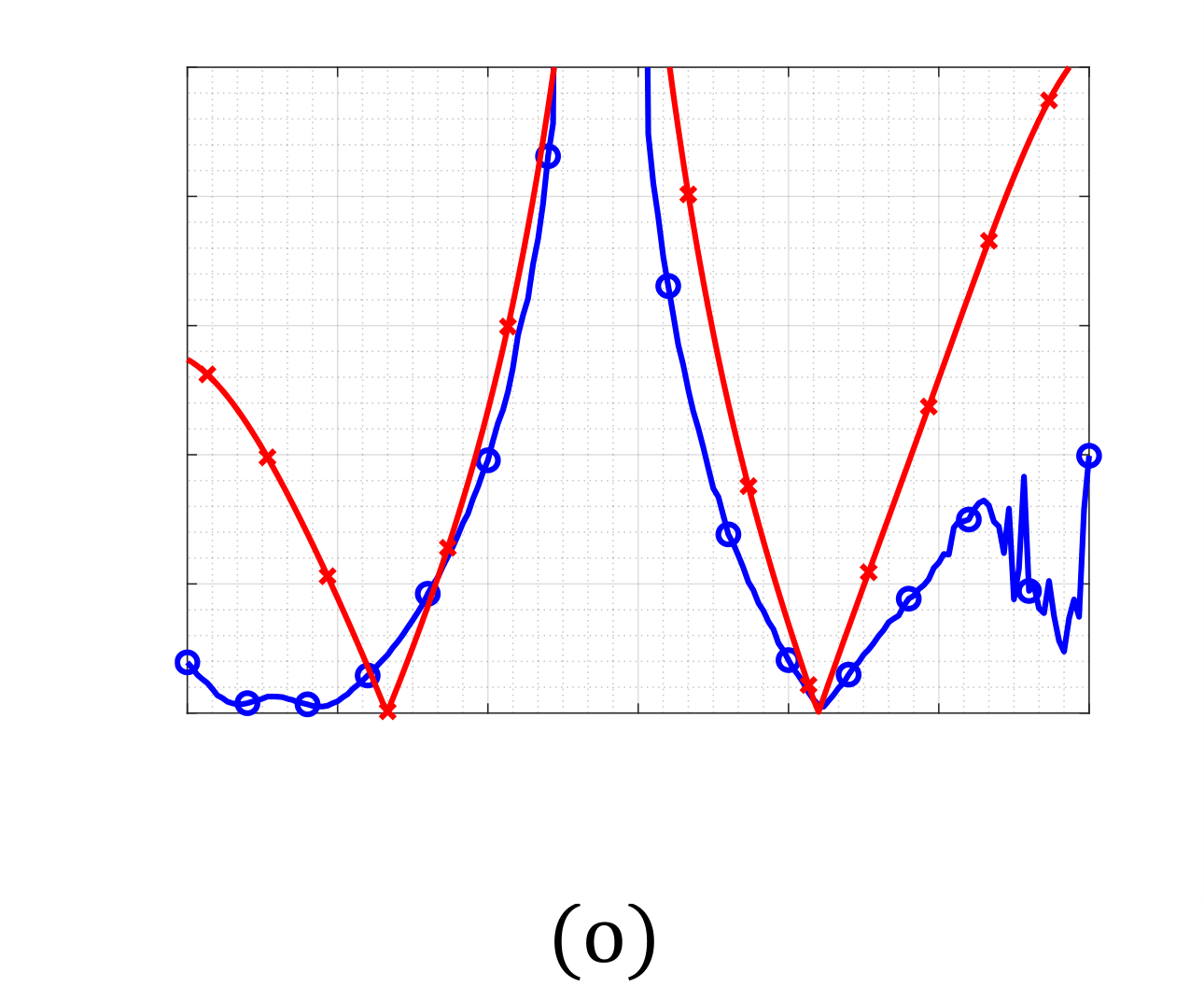}

\includegraphics[width=0.195\textwidth]{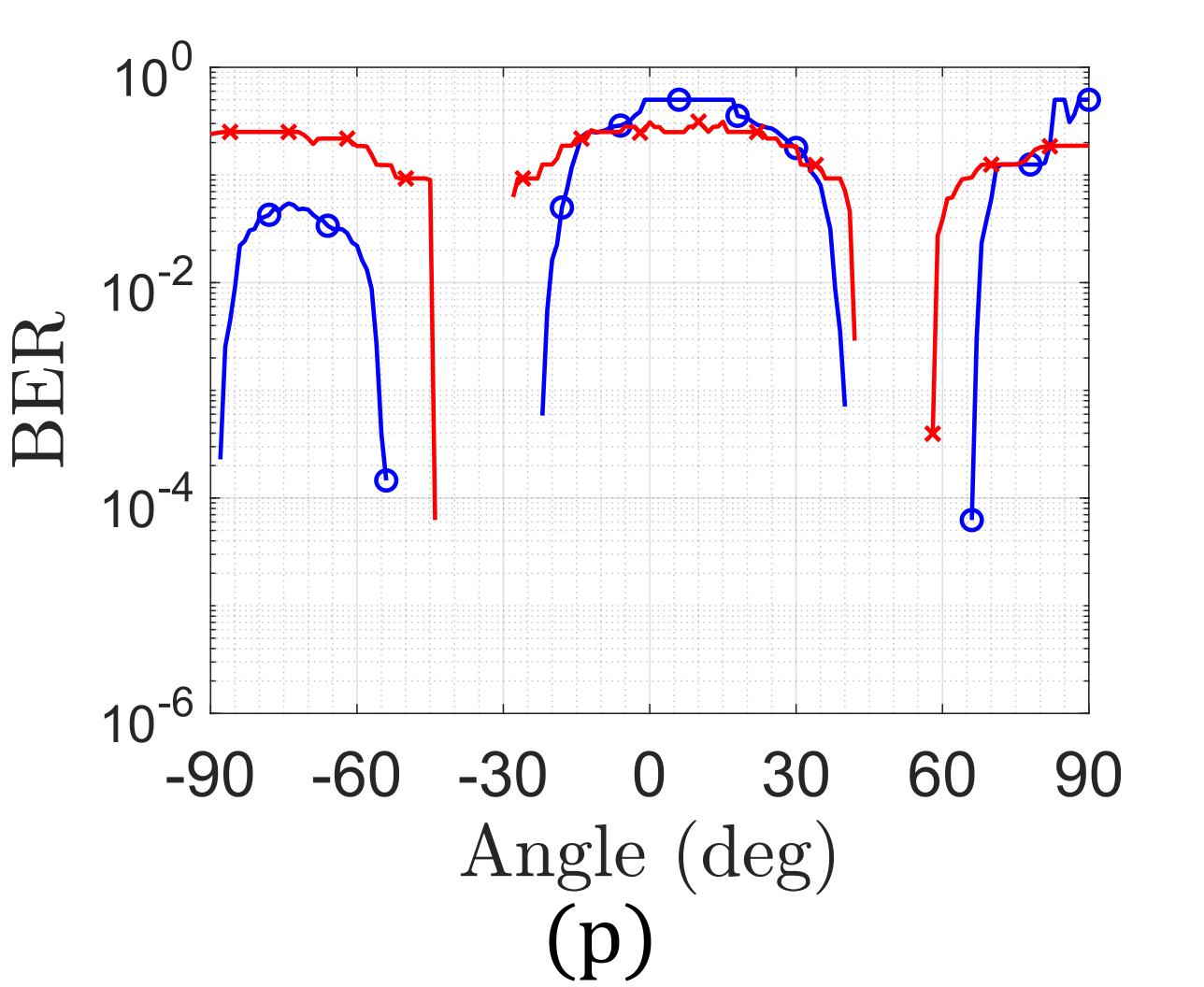}
\includegraphics[width=0.195\textwidth]{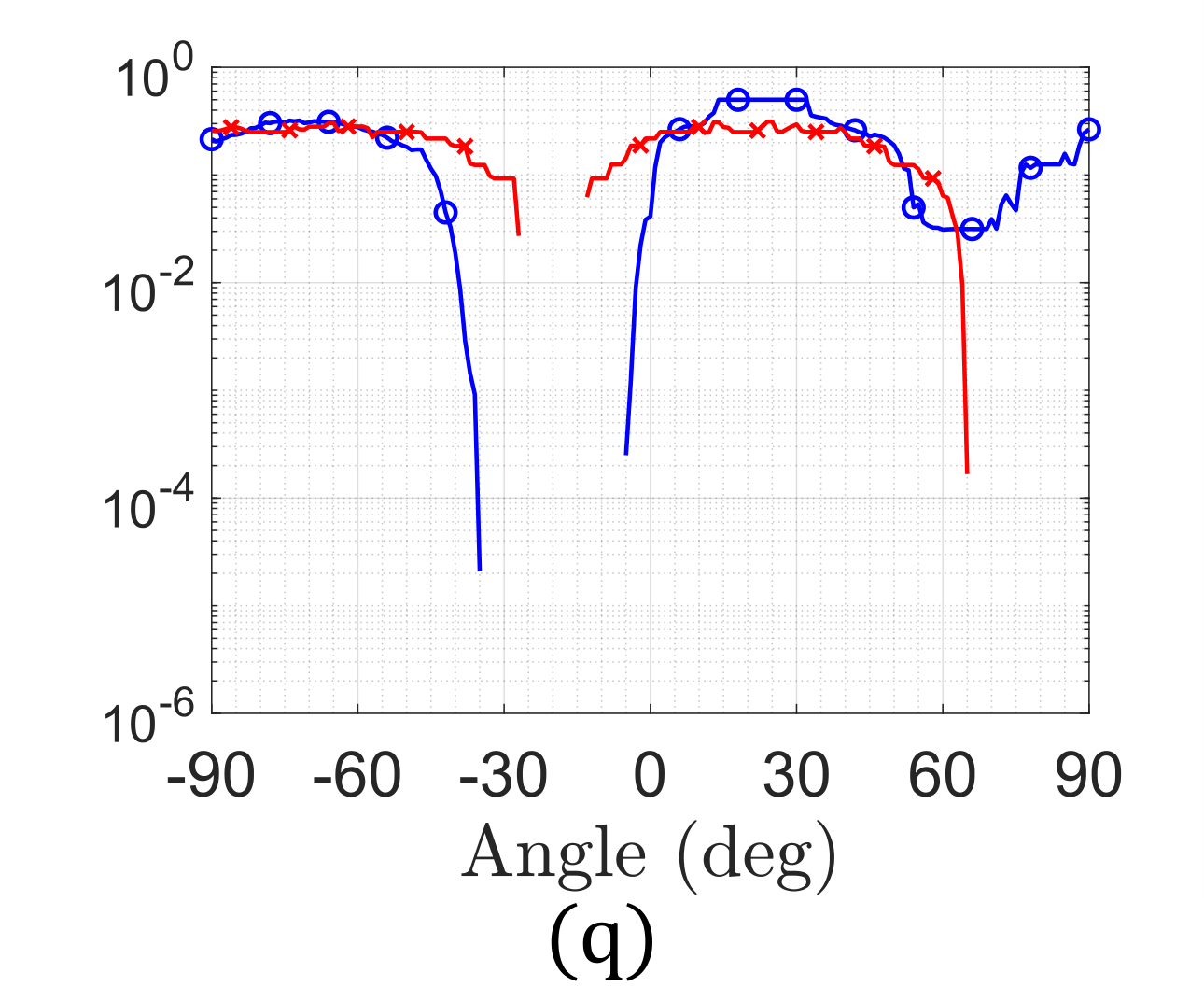}
\includegraphics[width=0.195\textwidth]{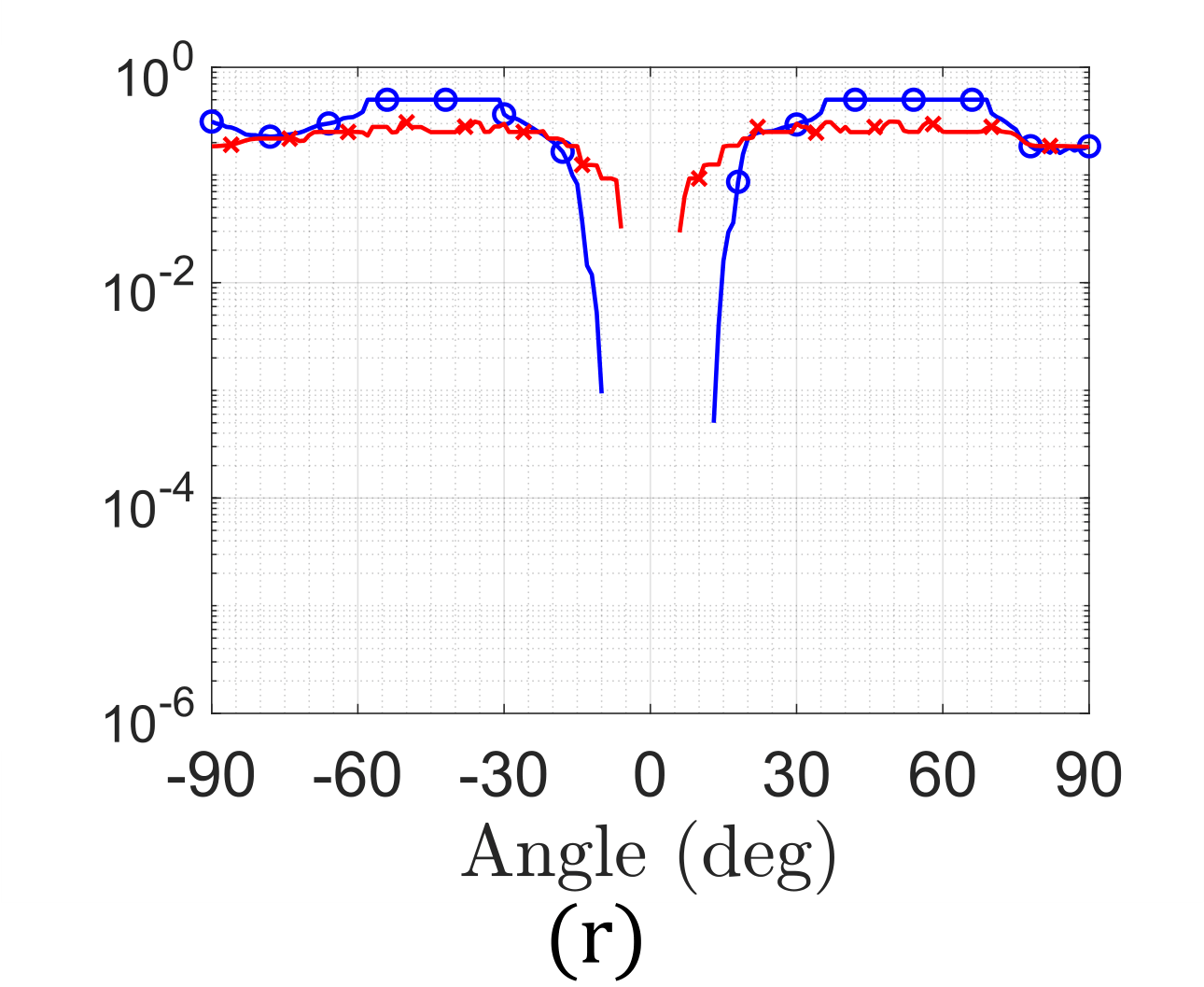}
\includegraphics[width=0.195\textwidth]{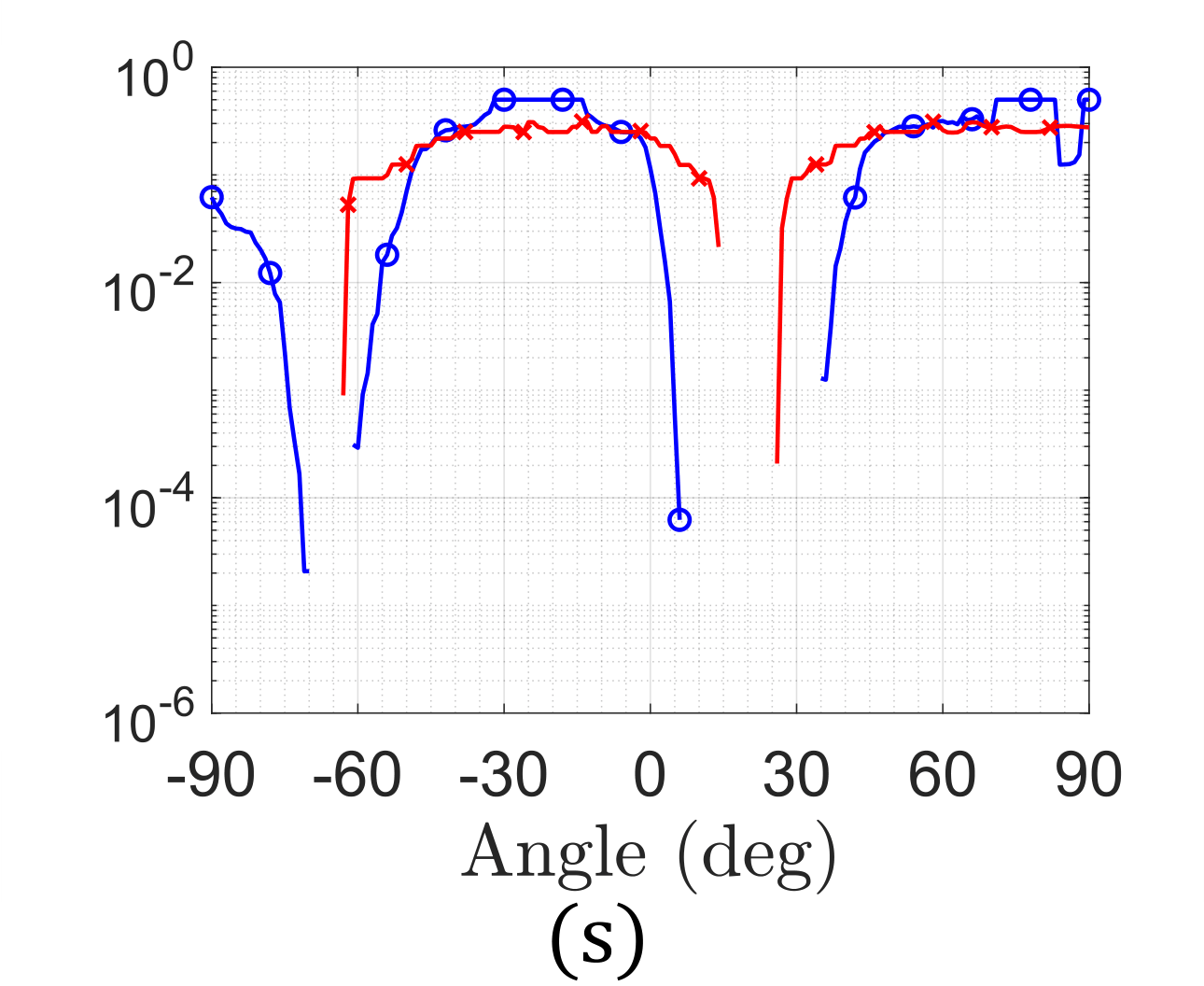}
\includegraphics[width=0.195\textwidth]{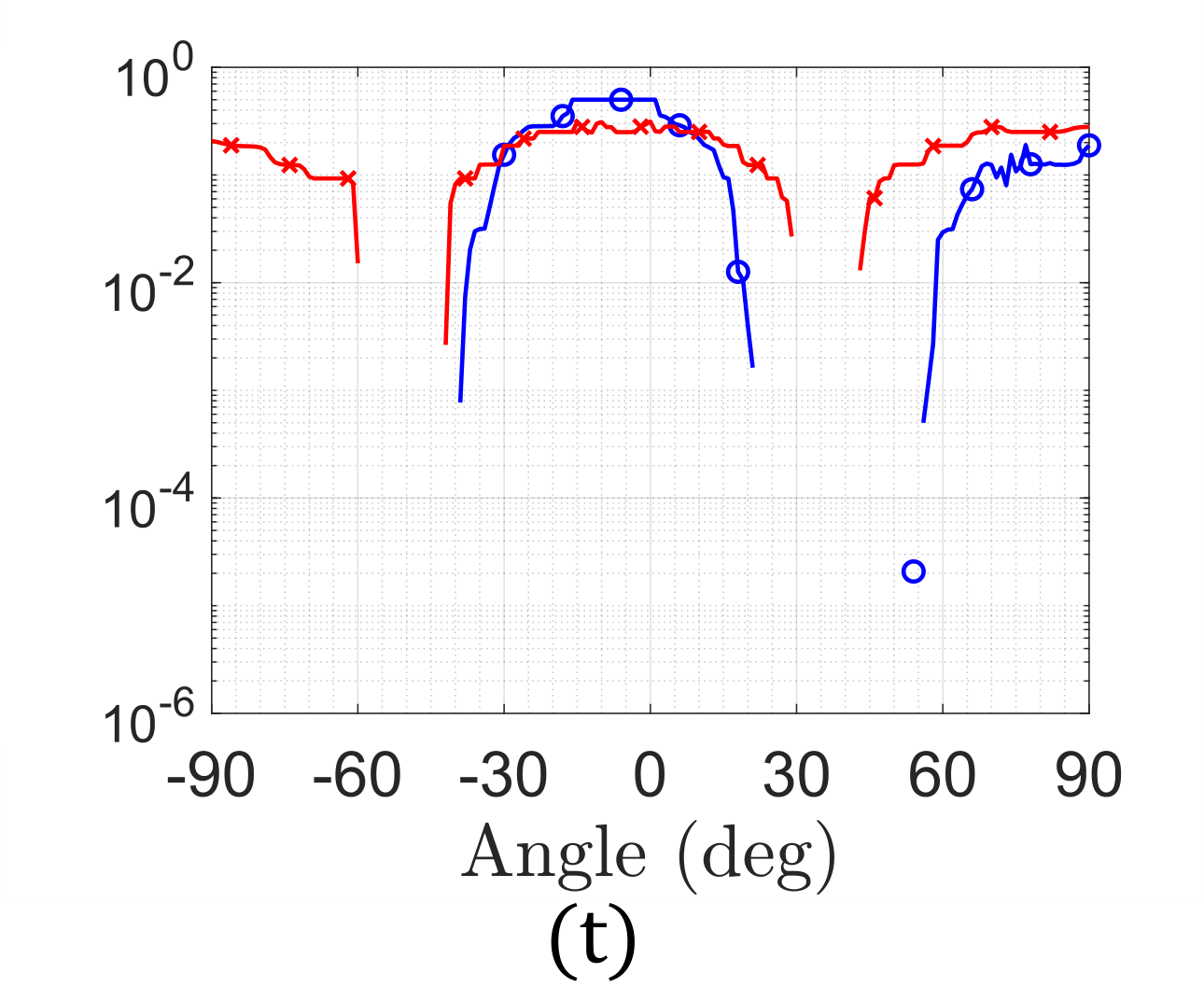}

\caption{The dynamic antenna phased array was steered using the communication system with a bandwidth of 1.5\;MHz switching at a rate of 3\;kHz with varying $\theta$ values to steer the information beam to (from left to right columns) $-35\degree$, $-20\degree$, $0\degree$, $20\degree$, and $35\degree$. The RMS power of the dynamic antenna phased array is shown in the top row (a-e), the RMS magnitude of the pattern difference between states 1 and 2 in the second row (f-j), the phase difference in the third row (k-o) and the BER in the fourth row (p-t). 
The information grating lobe that manifests when steering is due to the antenna element spacing of $0.75\lambda$ creating spatial ambiguities in the information beam.
}
\label{fig:Steerable_InformationBeam}
\end{figure*}

In this section we characterize the impact on the information beamwidth due to changing the switching rate of the dynamic antenna. We define the information beamwidth to be the region of space where BER < $10^{-3}$. We assume a 1 MHz symbol rate and experimentally evaluate switching rates of 
1\;kHz, 10\;kHz, 100\;kHz, 500\;kHz, 900\;kHz, 1.0\;MHz, 1.1\;MHz, and 2.0\;MHz. The measurements were conducted in the setup described in Fig.~\ref{fig:Measurement_Setup}(b), conducting measurements over $\phi\in[-90\degree, 90\degree]$ in $1\degree$ steps. Two receiver bandwidths were considered: a narrow bandwidth relative to the symbol rate of 1.5 MHz, and a wider bandwidth of 10 MHz. The results from the 1.5 MHz receiver bandwidth experiments are shown in Fig.~\ref{fig:Dynamic_Antenna_1500kHz_CommsSecurity} (a) which shows the average power spectral density over the signal bandwidth. 
The HFSS simulated power is computed by synthesizing the two-element array in HFSS and averaging the two states.
At low switching rates it can be seen that the measured response approximates the average of the two states, closely matching the simulation. As the switching frequency increases, the time-average pattern becomes more prominent, with nulls occurring at the expected angles of 
$\pm42\degree$. The implications on the communication security are then shown in the RMS magnitude error (the difference between the magnitudes of the two states as a function of angle), phase error (the difference between the phase of the two states), and bit error ratio (BER) in Figs.~\ref{fig:Dynamic_Antenna_1500kHz_CommsSecurity} (b), (c), and (d), respectively. As the switching frequency increases, fewer intermodulation products generated by the dynamic switching fall within the receiver bandwidth of 1.5\;MHz, reducing their impact on the BER. The information beam then widens as the switching frequency increases to a point where even at the 2.0\;MHz case, there are no intermodulation products captured by the receiver. 

With lower switching rates the magnitude and phase differences between the two states are larger at more angles, becoming less as the switching rate increases. In the spatial-spectral domain, more energy in the sidebands is filtered out by the narrow receiver bandwidth. The result is less modulation of the signals outside of the broadside direction, and in turn a wider region where the information is recoverable. The measured BER demonstrates this, as the region where the errors are minimal (corresponding to the information beam) widens as the switching rate increases. 



We evaluate the same range of switching frequencies with a wider receiver bandwidth of 10\;MHz to determine the impact of capturing more of the frequency sideband content generated by switching. The average power, relative errors in magnitude and phase between the two states, and the measured BER are shown as a function of angle in Fig.~\ref{fig:Dynamic_Antenna_8000kHz_CommsSecurity}.
The power is similar to the 1.5\;MHz case, however the regions where the magnitude and phase errors are small and thus the bit errors are low are considerably narrower than in the 1.5\;MHz case. The wider bandwidth receiver captures more of the information modulated outside the fundamental frequency band, contributing to greater errors. Note that on the BER curve there exist angles where less than 24,000 bits out of the total 48,000 bits were received by the communication system preventing accurate estimation of the errors, however in these cases the BER is inherently 0.5 or greater.  
The angle-frequency structure of the signals are shown in Fig.~\ref{fig:PSD_8000MHz} for the various switching rates. It can be seen that at lower switching rates the information is largely confined near the fundamental frequency, however as the switching rate increases the sideband structure becomes more prominent. At the most extreme, we see only one intermodulation product in Fig.~\ref{fig:PSD_8000MHz}(h).

\begin{table}
\caption{Information Beamwidth, BER < $10^{-3}$}
\centering
\begin{tabularx}{\columnwidth}{C|C|C}
\hline
Switching Frequency & 1.5\;MHz BW & 10\;MHz BW\\
\hline
1\;kHz   & 17\degree & 10\degree\\
10\;kHz  & 21\degree & 11\degree\\
100\;kHz & 26\degree & 20\degree\\
500\;kHz & 36\degree & 24\degree\\
900\;kHz & 54\degree & 29\degree\\
1.0\;MHz & N/A       & 35\degree\\
1.1\;MHz & N/A       & 41\degree\\
2.0\;MHz & N/A       & N/A\\
\hline
\end{tabularx}
\label{table:InformationBeam}
\end{table}

When considering the optimal switching rate, it is observed that switching slowly will result in the narrowest information beam. However, switching too slow will result in several symbols passing through the channel at a time and be appearing as static. The communication system in this work had a symbol rate of 1\;MHz and analyzed 1000 symbols at a time to estimate BER, resulting in the choice to switch at a minimum rate of 1\;kHz. The statistical probability that the antenna will switch during the 1000 symbols will decrease as the switching frequency goes below 1\;kHz. The BER of the dynamic phased array was measured in the same lab environment described in Fig. \ref{fig:Measurement_Setup}(b) and revealed that the frequency where the BER reduced 50\% when switching between 500 Hz to 900 Hz. It then collapsed to zero between 380\;Hz to 420\;Hz depending on the observation angle. Real-time magnitude changes were observed at angles with significant modulation; however, the EVM is minimal and the BER was zero for these low switching frequencies.

Experimental evaluation of the ability to steer the information beam was implemented at steering angles of $\phi~=~[-35\degree,\;-20\degree,\;0\degree,\;20\degree,\;35\degree]$ and compared to simulation in Fig.~\ref{fig:Steerable_InformationBeam}. Switching was implemented at 3\;kHz for all cases. The RMS magnitude error (second row) and phase error (third row) show good agreement to the simulated values. 
The BER (last row) shows a narrow region at the desired steering angles where information is transmitted with negligible errors.
When the information beam is steered to $\pm20\degree$, a second information beam appears centered at $\phi=\mp75\degree$, which is due to an information grating lobe that manifests because the element spacing is greater than $0.5\lambda$. The location of the information grating lobes can be calculated using~\eqref{eqn:FundamentalPower_MaxAngle}.
Note also that the information beam widens as the beam steers, similar to beam broadening away from broadside in a traditional phased array, both of which are due to the reduced projected area of the aperture. In the dynamic antenna cases, the reduced aperture results in the projection of the phase center motion to be reduced, which in turn broadens the information beam. These impacts are identical to beam steering and shaping impacts seen in radiated fields from traditional phased arrays, and therefore will have similar constraints and design considerations in terms of aperture size and element spacing.

\section{Conclusion}
We developed a general model for antenna array dynamics using phase center motion to directionally modulate an incoming or outgoing signal for a two-element phased array. 
We characterized the spatial-spectral response of the antenna in terms of a fundamental term and a term that captures the spatial distribution of energy in frequency sidebands. 
An experimental 2.5~GHz system was developed to evaluate the impact of switching rate and bandwidth, and to demonstrate beamsteering capability. Measurements matched closely to the model.
Results show that wireless communication security degrades as the switching frequency increases as a result of the receiver bandwidth filtering out the information modulated into the frequency sidebands. 
While switching slower provides a narrower information beam, the switching rate is lower bounded by the coherence time of the system such that the channel does not appear static outside the information beam. 

\bibliographystyle{IEEEtran}
\bibliography{bib}

\end{document}